\documentclass[12pt]{article}
\usepackage{amsmath}
\usepackage{amsfonts}
\usepackage{amssymb}
\usepackage{mathtools}
\usepackage{algorithm,algpseudocode}
\usepackage{amsthm}
\usepackage{natbib}
\usepackage{color}
\usepackage{caption}
\usepackage{subcaption}
\usepackage{graphicx}
\usepackage{pgfplots}
\pgfplotsset{compat=1.15}
\usepackage{tikz}
\usepackage{authblk}
\usepackage{multirow}
\usepackage{textalpha}
\usepackage{amsbsy}
\usepackage{graphics}
\usepackage{bm}
\usepackage{url}

\pgfplotsset{grid style={dashed,gray}}
\pgfplotsset{minor grid style={dotted,red!25}}
\usetikzlibrary{intersections}
\tikzset{help lines/.style={dashed, thick}}

\begin{document}

	\def\spacingset#1{\renewcommand{\baselinestretch}%
		{#1}\small\normalsize} \spacingset{1}

\title{Hypothesis Testing for Shapes using Vectorized Persistence Diagrams}
\author[1]{Chul Moon\thanks{Email: chulm@smu.edu}}
\author[2]{Nicole A. Lazar}
\affil[1]{Department of Statistical Science, Southern Methodist University, Dallas, TX, 75205, USA}
\affil[2]{Department of Statistics and Huck Institutes of the Life Sciences, Pennsylvania State University, University Park, PA, 16802, USA}

\date{}

\maketitle

\abstract{Topological data analysis involves the statistical characterization of the shape of data. Persistent homology is a primary tool of topological data analysis, which can be used to analyze topological features and perform statistical inference. In this paper, we present a two-stage hypothesis test for vectorized persistence diagrams. The first stage filters vector elements in the vectorized persistence diagrams to enhance the power of the test. The second stage consists of multiple hypothesis tests, with false positives controlled by false discovery rates. We demonstrate the flexibility of our method by applying it to a variety of simulated and real-world data types. Our results show that the proposed hypothesis test enables accurate and informative inferences on the shape of data compared to the existing hypothesis testing methods for persistent homology.}

\noindent%
	\vfill
	
	\newpage
 \spacingset{1.45} 


\section{Introduction}
\begin{figure}[!ht]
		\centering
		\subcaptionbox{Bentheimer}[.2\textwidth]{
			\includegraphics[width=1\linewidth]{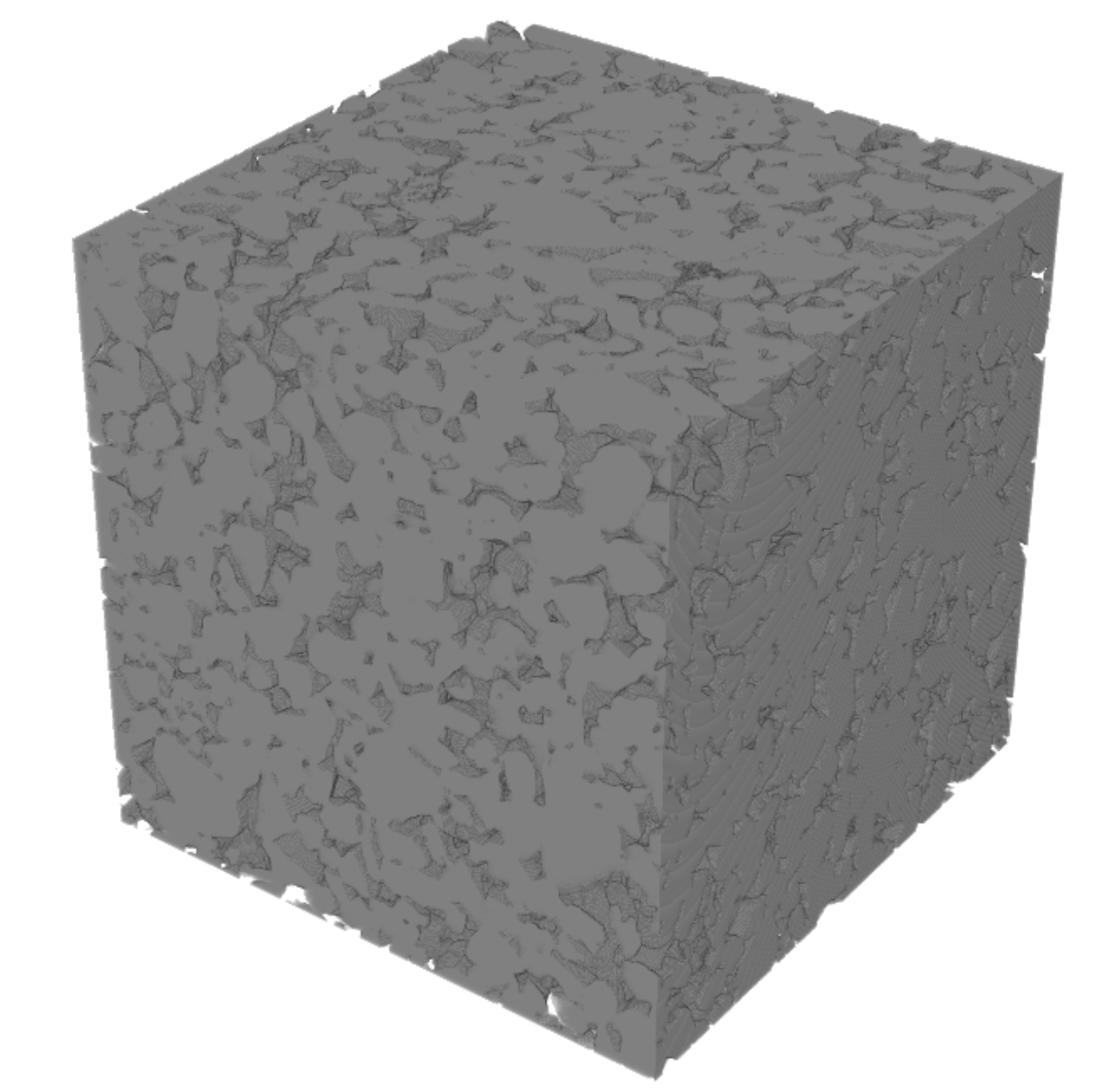}
		}
		\subcaptionbox{Doddington}[.2\textwidth]{
			\includegraphics[width=1\linewidth]{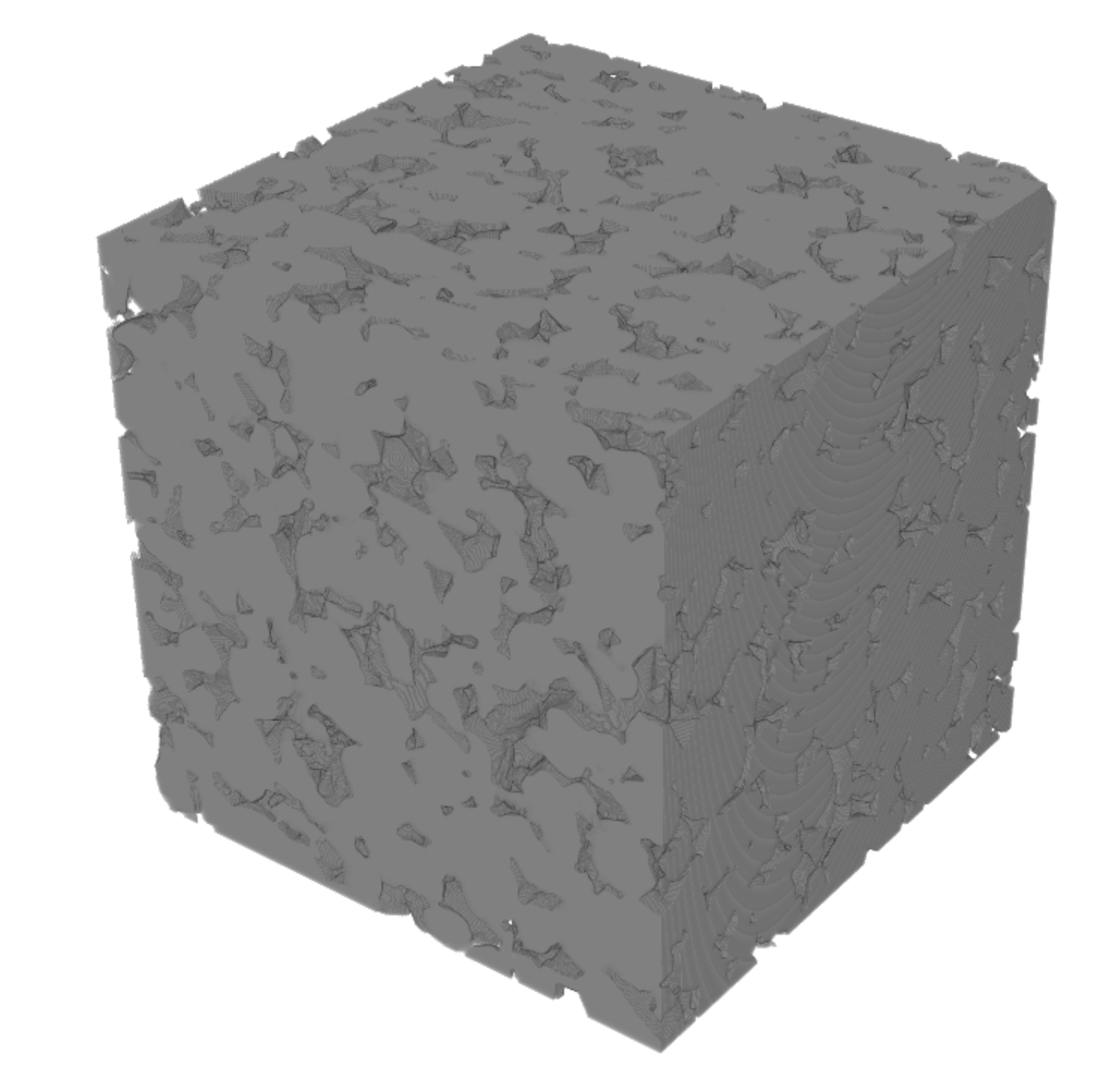}
		}
		\subcaptionbox{Estaillades}[.2\textwidth]{
			\includegraphics[width=1\linewidth]{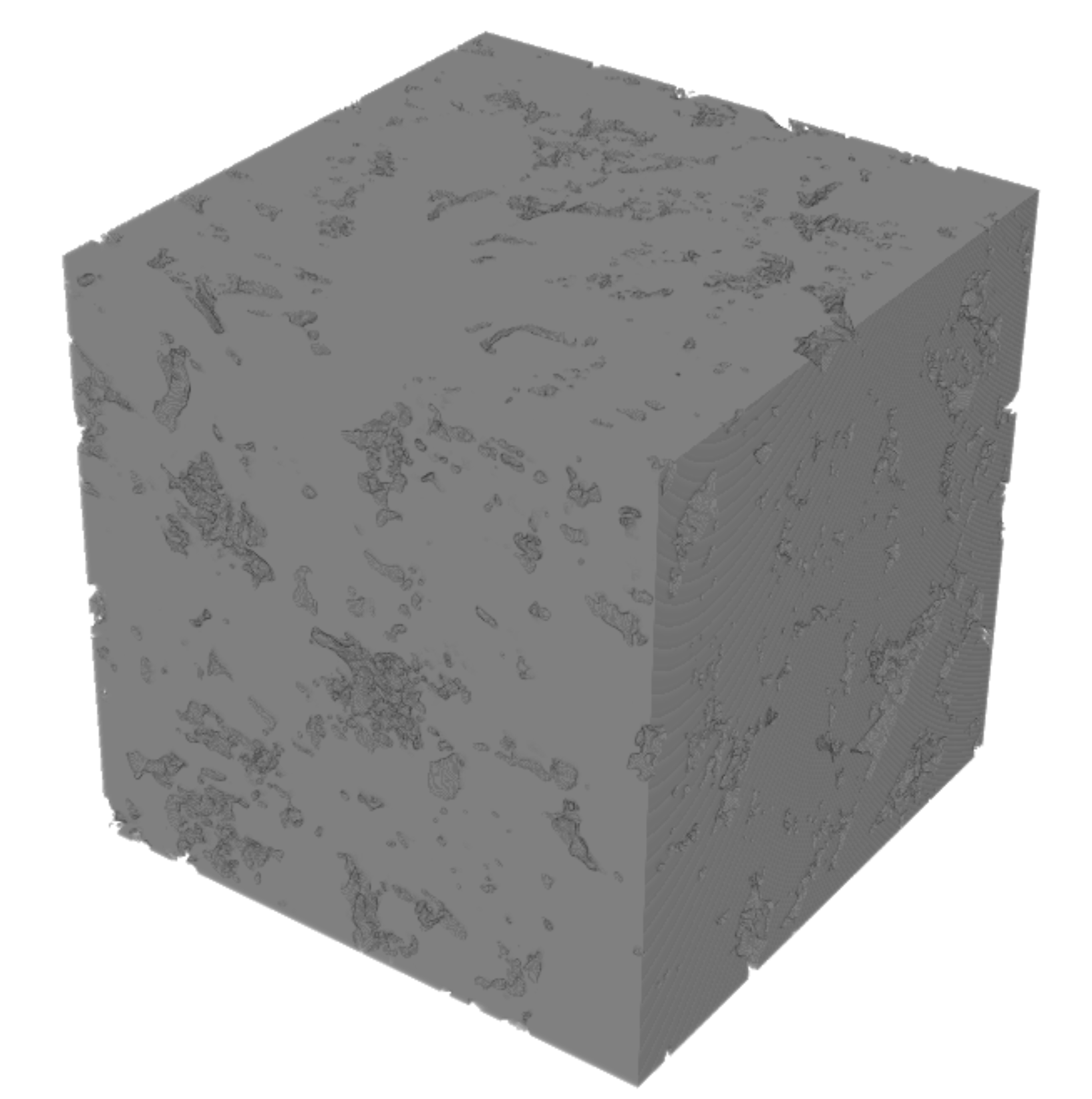}
		}
		\subcaptionbox{Ketton}[.2\textwidth]{
			\includegraphics[width=1\linewidth]{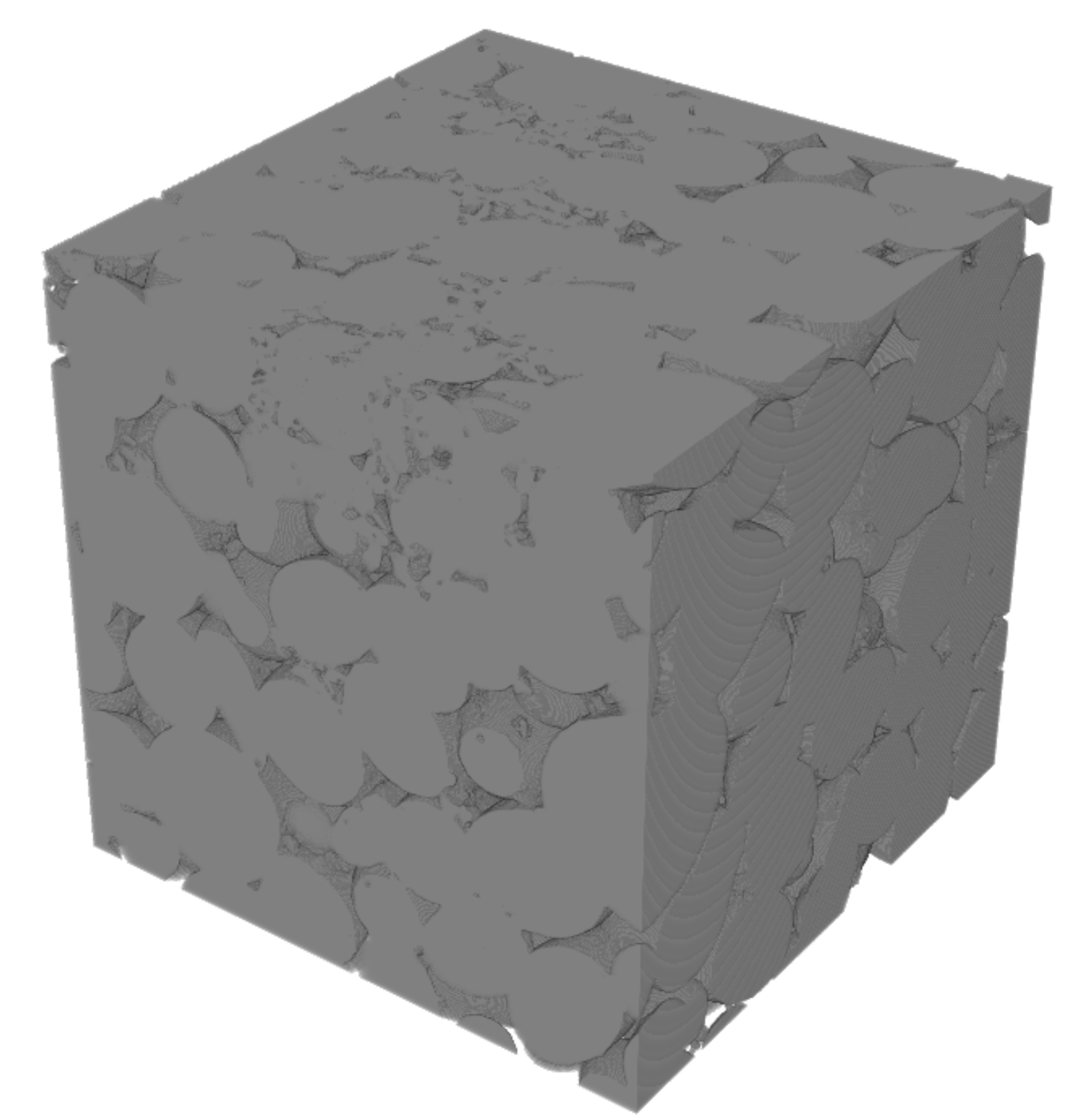}
		}
		\caption{Micro-CT images of Bentheimer and Doddington sandstones and Estaillades and Ketton carbonate rocks \citep{icl}.}
		\label{fig:rocks}
	\end{figure}

Modern science is facing a rapid increase in the volume of data as well as their complexity. Non-standard data types such as functional \citep{Ramsey2005}, manifold \citep{Genovese2012}, and object-oriented \citep{Marron2014} data have become more common.
	New methodologies need to be developed to analyze and gain useful information from these data. 
	For example, Figure~\ref{fig:rocks} shows Micro-CT images obtained from four different types of rocks. 
	The properties of rocks, such as permeability, play an important role in earth science fields, including hydrogeology and petroleum engineering. 
	However, such properties are difficult to estimate; one of the reasons is that they are influenced by parameters linked to pore geometry \citep{Hommel2018}.
	Because it is not easy to quantify pore geometry, various indirect measures are used instead, such as porosity, tortuosity, and specific surface area \citep{Bernabe1998,XIONG2016}.

Topological data analysis (TDA) is a recent addition to the analytical toolbox, that quantifies the shape of data using their topological features, such as connected components and holes. 
	A primary TDA tool is persistent homology, which analyzes topological features of data in various scales \citep{Edelsbrunner2008,Carlsson2009}.
	Persistent homology provides a numeric descriptor of the shape of data that is robust to noise and insensitive to metrics \citep{Chazal2017}.
	It has been applied to a wide variety of data sets, including atomic configurations \citep{Nakamura2015,Hiraoka2016}, biomolecular structures \citep{kovacev2016using,Cang2017,sauerwald2019topological,Townsend2020}, brain arteries \citep{Bendich2016}, financial time series data \citep{aguilar2020topology}, metalic materials \citep{kimura2018}, tumor images \citep{Lawson2019,moon2020using}, and porous materials \citep{Robins2016,Jiang2018,Herring2019,Moon2019}. These applications have shown that topological features successfully characterize the shape of data.

 Using results from TDA, it is possible to move toward statistical inference on the shape of data. 
	For example, by computing persistent homology of the four rocks in Figure~\ref{fig:rocks}, we can obtain numeric outputs that describe their shapes.
	We can then, in principle, use this output to conduct hypothesis tests to distinguish rock types based on characteristics such as pore shapes and connectivities.  
	
	Most statistical and machine learning methods, however, cannot be directly applied to persistent homology results. Outputs of persistent homology are algebraic objects, not vectors. The numeric summary of persistent homology outputs are sets of intervals that describe how topological features persist. 
	Therefore, we cannot simply extend approaches developed for symbolic data, such as \cite{gioia2005basic} and \cite{Billard2006}, to persistent homology results.
	Various approaches have been suggested to represent persistent homology results in different spaces while preserving the summarized topological information; in Euclidean space \citep{Bendich2016tracking, Adcock2016, Adams2017,Kalisnik2018},
	reproducing kernel Hilbert space (RKHS) \citep{Reininghaus2015,kusano2017kernel},
	and $L^2$-space \citep{Bubenik2015}.

	Several methods have been proposed to make statistical inference using persistent homology results including
	confidence intervals using bootstrap \citep{Fasy2014}, linear models \citep{Obayashi2018}, and
	Bayesian approaches \citep{Maroulas2019}. 
	There are a few authors who describe hypothesis testing procedures for persistent homology: permutation tests using a pairwise distance of persistence diagrams \citep{Robinson2017,cericola2018} and 
	functional summaries \citep{Chen2015,Berry2020}, multiple hypothesis testing procedures using uniform point cloud data \citep{Vejdemo2018}, tests using persistence landscapes \citep{Bubenik2015}, and kernel two-sample hypothesis tests in RKHS \citep{kwitt2015statistical,Kusano2019}. However, the existing approaches provide results of limited interpretability. 
	
	In this paper, we propose a two-stage hypothesis test of filtering and testing for the persistence image suggested by \cite{Adams2017} that represents persistent homology features as vectors in Euclidean space.
	The two-stage hypothesis test has been used to enhance detection power for various high-dimensional data such as microarray \citep{Hackstadt2009,Tritchler2009} and genome-wide association analyses \citep{Murcray2008,Kooperberg2008}. 
	The filtering step removes pixels in the persistence image that may not be relevant to the inference.
	The testing stage performs multiple hypothesis tests and controls the false discovery rate (FDR).
	The proposed method is flexible in the sense that it can handle a wide variety of data types.

	The rest of the paper is organized as follows. 
	In Section~\ref{sec:background}, we provide background on persistent homology, topological features for tomographic image data, and issues related to hypothesis tests for vectorized persistence diagrams. 
	Section~\ref{sec:ts} introduces the proposed two-stage hypothesis test procedures. 
	Section~\ref{sec:sim} presents the hypothesis testing results for simulated point clouds, pseudo-material images, and beetle population data.  
	Section~\ref{sec:app} applies the proposed method to sand pack images and musical instrument sounds.
	Finally, in Section~\ref{sec:discussion}, we discuss our main contributions and future directions.

	\section{Topological Data Analysis Background}
	\label{sec:background}

	\subsection{Homology and Persistent Homology}

	Homology characterizes a shape by counting its number of connected components, loops, and voids. 
	The connected components, loops, and voids are also called the dimension-zero, dimension-one, and dimension-two topological features, respectively. 
	Persistent homology tracks the dynamics of topological features in data at different resolutions. As we change the resolution of the data, the topological nature of the data also changes. The parameter that tracks the resolution of data is called the filtration.
	Persistent homology computes when specific dimension-$k$ features appear (birth) and disappear (death) over the filtration.
	The output of persistent homology can be summarized as a collection of intervals of (birth, death). 
	
\begin{figure*}[!t]
		\centering
		\begin{subfigure}{0.32\textwidth}
			\centering
			\includegraphics[height=0.8\linewidth]{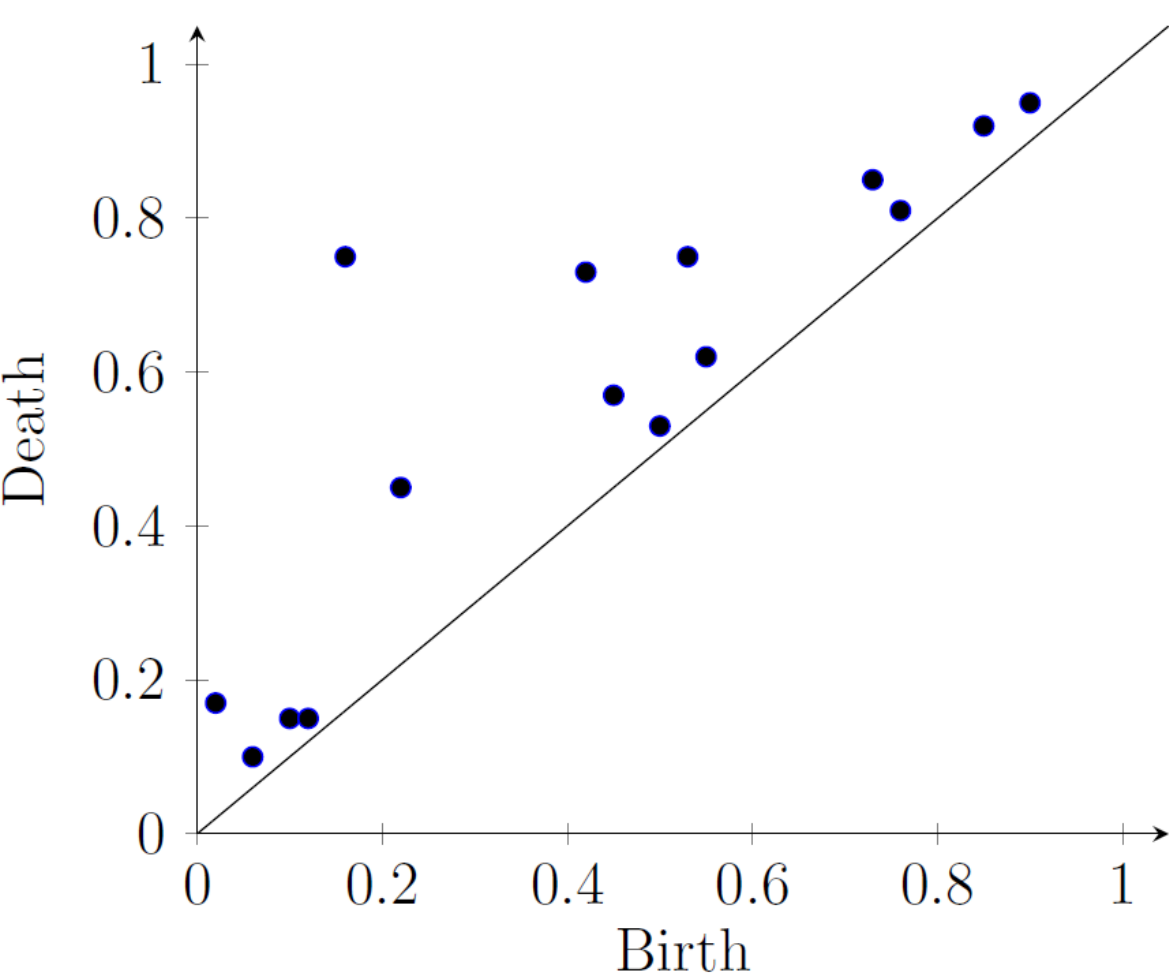}
			\caption{Persistence diagram}\label{subfig:pd}
		\end{subfigure}
		\begin{subfigure}{0.32\textwidth}
			\centering
			\includegraphics[height=0.8\linewidth]{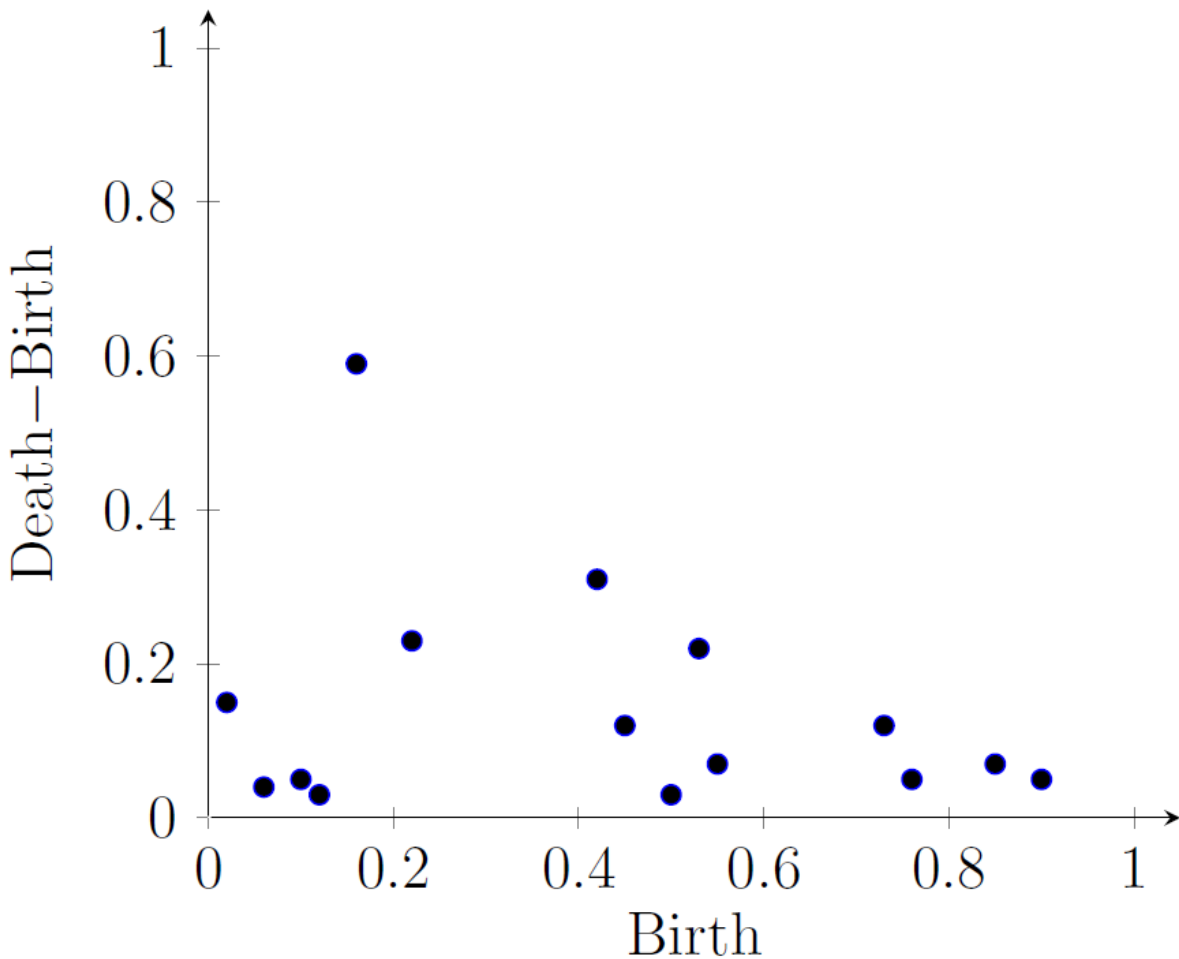}
			\caption{Transformed persistence diagram}\label{subfig:tpd}
		\end{subfigure}
		\begin{subfigure}{0.32\textwidth}
			\centering
			\includegraphics[height=0.8\linewidth]{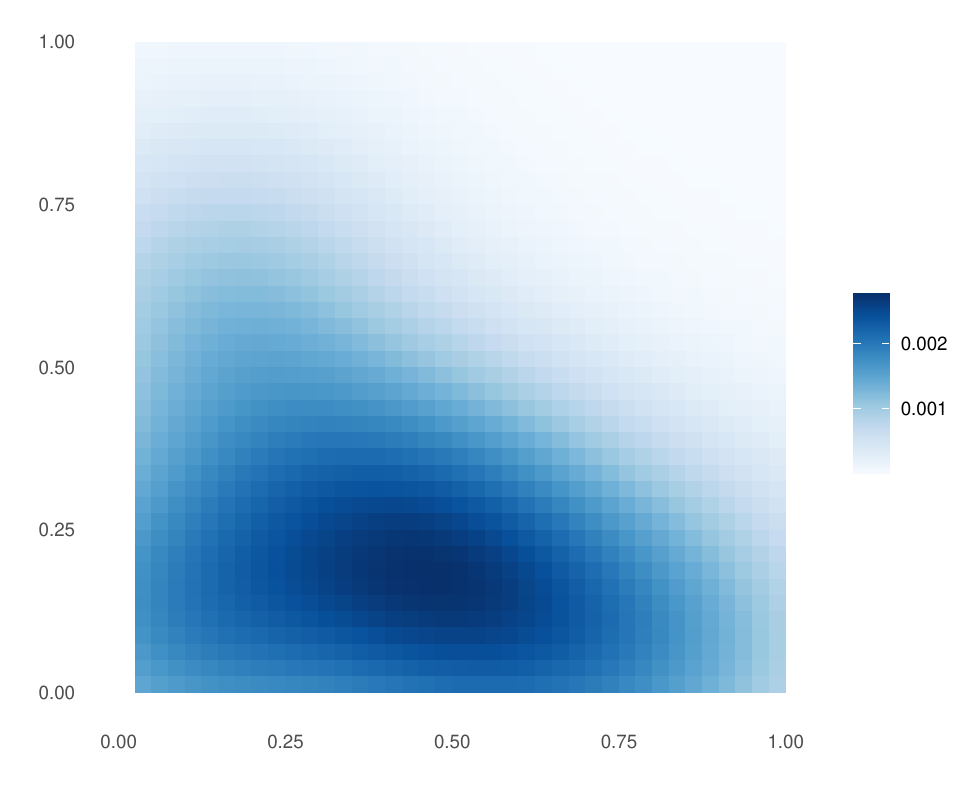}
			\caption{Persistence image}\label{subfig:pi}
		\end{subfigure}
		\caption{Steps of converting the persistence diagram into the persistence image. From the original persistence diagram (left) to the transformed persistence diagram (middle) to the persistence image (right). In this example, the Gaussian smoothing function and the arctangent weight are used for the persistence images.}
		\label{fig:PI}
	\end{figure*}

	A persistence diagram is the most popular graphical representation of the (birth, death) intervals produced by persistent homology. It plots the intervals as points in $\mathbb{R}^2$ of birth ($x$-axis) and death ($y$-axis). Because the death of a topological feature comes after its birth, all points are plotted above the diagonal. Figure~\ref{subfig:pd} shows an example of a persistence diagram. 
	For more detailed explanation of persistent homology, see Section~S2 of the supplementary material.

	\subsection{Representations of Persistence Diagrams}
	
	Although persistence diagrams include information about topological features, it is difficult to use them as input in data analysis.
	This is because persistence diagrams are not numeric vectors, which most classical statistical and machine learning methods require. 
	Ways to represent persistence diagrams as vectors include binning \citep{Bendich2016tracking}, polynomial \citep{Adcock2016}, persistence image \citep{Adams2017}, tropical polynomial \citep{Kalisnik2018} in Euclidean space, persistence scale-space kernel \citep{Reininghaus2015} and persistence weighted Gaussian kernel \citep{kusano2017kernel} in RKHS, and rank function \citep{Robins2016rank} and persistence landscape \citep{Bubenik2015} in $L^2$ space. 
	Every representation has its advantages and shortcomings.
	For example, the polynomials by \cite{Adcock2016} are easy to compute, but not Lipschitz continuous. Also, tropical polynomials by \cite{Kalisnik2018} generate sufficient statistics for topological summaries \citep{monod2019tropical}, but are not composed of a well-defined and rigorous ring structure. 
	
	We develop a hypothesis test procedure for persistence diagrams represented in Euclidean space, specifically using the persistence images. Persistence images have two main advantages. First, the representation in Euclidean space makes it easier to conduct prediction and classification by applying classical statistical models and machine learning methods such as regression and support vector machine \citep{Adams2017, Obayashi2018}. Second, they have a natural graphical summary unlike most other representations.

	The persistence images are obtained by the following steps. First, a persistence diagram $\text{PD}=\{(\text{birth},\text{death})\}$ is transformed into $\text{PD}_t=\{(u=\text{birth},v=\text{death}-\text{birth})\}$. With a real-valued smoothing function $f_{(u,v)}$ for $(u,v) \in \text{PD}_t$ and a real-valued weight function $w$, the persistence surface $\rho_{\text{PD}_t}$ of the transformed persistence diagram $\text{PD}_t$ is 
	\begin{equation*}
	\rho_{\text{PD}_t} (x,y) = \sum\limits_{(u,v) \in \text{PD}_t} f_{(u,v)}(x,y) \cdot w(u,v),
	\end{equation*}
	where $x$ and $y$ are the $(x,y)$-coordinates of the persistence surface.
	The persistence images are obtained by taking an integral of $\rho_{\text{PD}_t}$ over a given grid. 
	We use the Gaussian smoothing function $f_{(u,v)}(x,y \mid h) = \frac{1}{2\pi h^2 }\exp( -( (x-u)^2 + (y-v)^2 ) /2h^2)$. 
	The values of $h$ can be determined case by case, but the persistence image is known to be robust to the choice of $h$ \citep{Adams2017}.
	\cite{Adams2017} show that the persistence image is stable with respect to bottleneck and Wasserstein distances  \citep{Edelsbrunner2010}.

	The weight function in persistence images can highlight specific aspects of topological features.
	A suitable weight can reveal key features about the shape of data and various weights have been proposed.
	For example, a linear weight ($w(u,v)=v$) assigns higher weights to persistent topological features whereas the constant weight ($w(u,v)=1$) gives equal weights to all features. Also, an arctangent weight ($w(u,v)=\arctan( Rv^{S})$, for $R,S\in \mathbb{R}$) is suggested in \cite{kusano2017kernel} for RKHS and extended to persistence images in \cite{Obayashi2018}. In our study, we use the arctangent weight and set $R=S=0.5$. Note that the weights used in our study depend only on the $v=\text{death}-\text{birth}$, the persistence of the feature.

	Figure~\ref{fig:PI} illustrates the conversion steps from the persistence diagram to the persistence image. First, the (birth, death) pairs of the persistence diagram in Figure~\ref{subfig:pd} are transformed to the (birth, death$-$birth) pairs in Figure~\ref{subfig:tpd}. Then, the transformed persistence diagram is represented as the persistence image in Figure~\ref{subfig:pi}. Here, the Gaussian smoothing function and the arctangent weight are used.

	\subsection{Hypothesis Tests for Persistent Homology}
	One can consider a hypothesis test that compares the topological features of data using persistence diagrams.
	However, it is difficult to define probability distributions on the space of persistence diagrams because the space is infinite in dimension and has complicated geometry \citep{Robinson2017}. 	
	As a result, the hypothesis test methods for persistent homology have been studied in two ways: 1) using permutation tests and 2) using statistical properties of represented persistence diagrams. 
	
	First, the permutation-based tests are defined on the space induced by pairwise metrics of persistence diagrams or their representations.
	\cite{Robinson2017} propose a permutation test for a two-sample setting using persistence diagrams. 
	Assume that we have $n=n_1+n_2$ persistence diagrams $\text{PD}_l$, $l\in\{1,2,\cdots,n\}$, obtained from two data groups of size $n_1$ and $n_2$. 
	Let $G=\{G_1,G_2\}$ be labels for the two groups where $G_1,G_2 \subset \{1, 2, \cdots, n \}$, $G_1 \cap G_2 = \emptyset$, $G_1 \cup G_2 = \{1, 2, \cdots, n \}$, $n(G_1)=n_1$, and $n(G_2)=n_2$.  
	The permutation test uses a loss function $L$ of a label $G$ defined by 
	$
	L(G=\{G_1,G_2\})= \frac{2}{n_1(n_1-1)} \sum_{l_i<l_j \in G_1} d(\text{PD}_{l_i},\text{PD}_{l_j}) + \frac{2}{n_2(n_2-1)}\sum_{l_i<l_j \in G_2} d(\text{PD}_{l_i},\text{PD}_{l_j}),
	$
	where $d$ is a pairwise distance function for persistence diagrams such as bottleneck or Wasserstein distance. The loss function measures the similarity of persistence diagrams within groups: if persistence diagrams within groups are similar to each other, the loss will be small, and vice versa. At first, the baseline loss value $L_0(G_0)$ is computed for the initial group label $G_0$ that we want to test. Then the labels are randomly shuffled and the loss values are computed using the permuted labels. The permutations are repeated $N_P$ times. The p-value is computed by comparing the baseline loss value $L_0(G_0)$ with the $N_P$ loss values of the permuted labels. Because the number of permutations $N_P$ can be up to $\binom{n}{n_1}$, a smaller number of permutations can be used instead.
	Algorithm~S1 in the supplementary Material summarizes the permutation test procedure of \cite{Robinson2017}. 
	\cite{cericola2018} extend the two-sample test scheme to multiple label group testing using the one-way analysis of variance (ANOVA) procedure and \cite{Vejdemo2018} propose procedures to control the multiple testing problem.
	The permutation test is also suggested for the persistence landscape \citep{Bubenik2015}. For the persistence landscape, the loss function $L(G=\{G_1,G_2\})$ is defined by the distance between the mean persistence landscapes of the two groups.
	Also, \cite{Chen2015} and \cite{Berry2020} propose permutation tests using functional representations of persistence diagrams.

	Second, hypothesis tests are conducted on the various persistence diagram representations.
	\cite{Bubenik2015} suggests the two-sample z-test for mean persistence landscapes and Hotelling's $T^2$ test for vectors of persistence landscape functionals. However, these tests may be less powerful than the permutation test in some cases \citep{Bubenik2015}. 
	Also, \cite{kwitt2015statistical} and \cite{Kusano2019} apply the kernel two-sample test of \cite{Gretton2006,Gretton2012} to persistence diagram representations in RKHS. 
	The kernel two-sample test uses the maximum mean discrepancy (MMD) as a test statistic that measures the difference between the kernel functions. In the kernel test, the MMD of the two groups is computed and it is compared with the $1-\alpha$ quantile of the null distribution of the MMD. 
	
\begin{table*}[t]
	\caption{Comparison of hypothesis testing methods for persistent homology.\label{tb:method}}
	\tabcolsep=0pt
\begin{tabular*}{\textwidth}{@{\extracolsep{\fill}}ccccc@{\extracolsep{\fill}}}
\hline
Representation        & Testing method   & Weight     & Testing result   & Result visualization \\ \hline
Persistence image     & Proposed two-stage test & Flexible   & Multiple p-values & Yes           \\ 
Persistence diagram   & Permutation      & Inflexible & Single p-value   & No            \\ 
Persistence landscape & Permutation      & Inflexible & Single p-value   & No            \\ 
   & $z$-test \& Hotelling's $T^2$ test       & Inflexible & Single p-value   & No            \\ 
RKHS vector           & Kernel test      & Flexible   & Single p-value   & No            \\ \hline
\end{tabular*}
\end{table*}

	Existing hypothesis test approaches proposed for persistent homology have some limitations. Table~\ref{tb:method} compares the existing methods. First, the existing hypothesis test methods provide limited information on how each topological feature contributes to any observed differences. They provide a single p-value as a testing result.
	Therefore, it is difficult to identify which topological features play an essential role in the hypothesis test. 
	Second, most methods do not provide an option for assigning weights to topological features that enables flexible interpretation. For example, the hypothesis tests using persistence diagrams and persistence landscapes are only available under the fixed weight. We note that other persistent homology representations such as weighted silhouette \citep{chazal2014stochastic} are not compared in Table~\ref{tb:method} because their hypothesis testing methods have not been studied.

	Conducting hypothesis tests on persistence images can overcome the limitations; it can identify topological features that account for the differences, visualize them, and implement weights.
	However, a naive application of hypothesis tests to persistence images can lead to an incorrect conclusion.

	\subsection{Naive Hypothesis Test for Persistence Images}
	\label{subsec:tstt}
	
	For the vectorized persistence diagrams in Euclidean space, we can measure the differences by comparing the mean values of vectors. 
	We consider testing the difference between two groups of spaces ($\mathcal{S}_1$ and $\mathcal{S}_2$) for simplicity. It is straightforward to generalize the approach to more than two groups, similar to the transition from t-tests to ANOVA.
	Assume that $n_1$ and $n_2$ persistence diagrams are obtained from $\mathcal{S}_1$ and $\mathcal{S}_2$, respectively, and represented as persistence images of $m\times m$ pixels. 
	Let $x_{(j,k)}^{i}$ be the value of the $i$th pixel of group $j$'s $k$th persistence image, where  $i\in \{ 1, \cdots, m^2 \}$, $j=1,2$, and $k\in \{ 1, \cdots, n_j \}$. Also, let $\mu_j^i=E_{k}(x^i_{(j,k)})$ be the mean of the $i$th pixel of group $j$'s persistence images. 
	Then, the hypotheses for testing differences between two persistence images are
	\begin{eqnarray*}
		H_0:&& \mu_1^{i} = \mu_2^{i} \text{ for all } i \in \{ 1, \cdots, m^2\} \nonumber \\
		H_1:&& \mu_1^{i} \neq \mu_2^{i} \text{ for at least one } i \in \{ 1, \cdots, m^2\}.
		\label{eqn:hypo}
	\end{eqnarray*}	
	
	However, it is not appropriate to implement these naive hypotheses directly. First, not all areas of the persistence images are equally interesting. There will most likely be multiple pixels with a mean close to zero in both groups just because they are far from the diagonal or are in otherwise sparse regions.
	Second, the pixels are not independent. The persistence image uses a smoothing function, so nearby pixels become dependent.
	
		\begin{figure}[!t]
		\centering
		\begin{subfigure}{0.23\textwidth}
			\centering
			\includegraphics[height=1.22\linewidth]{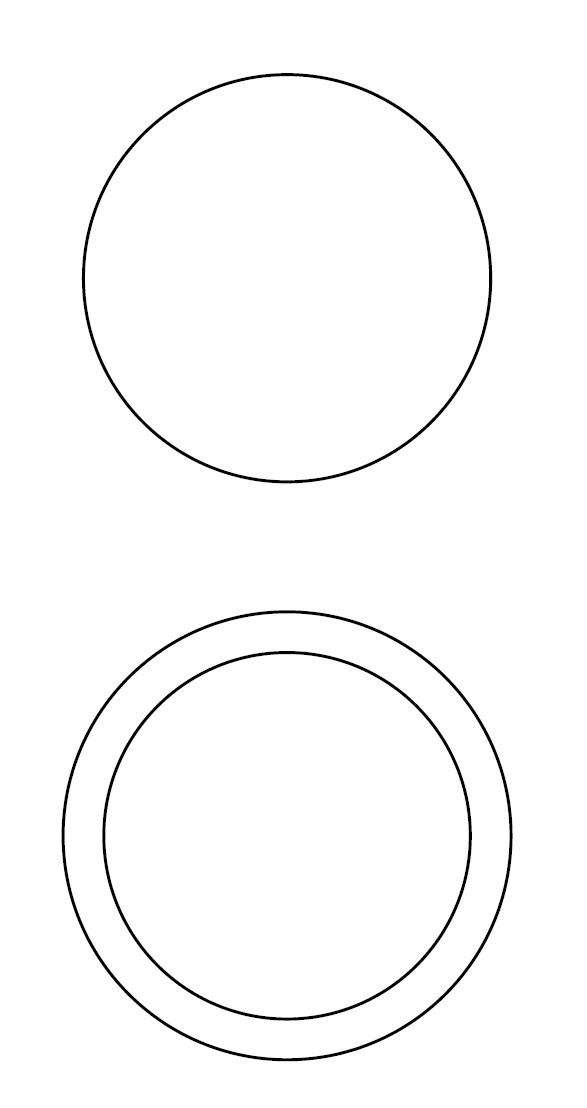}
			\caption{Shapes}\label{subfig:shape}
		\end{subfigure}
		\begin{subfigure}{0.23\textwidth}
			\centering
			\includegraphics[height=1.22\linewidth]{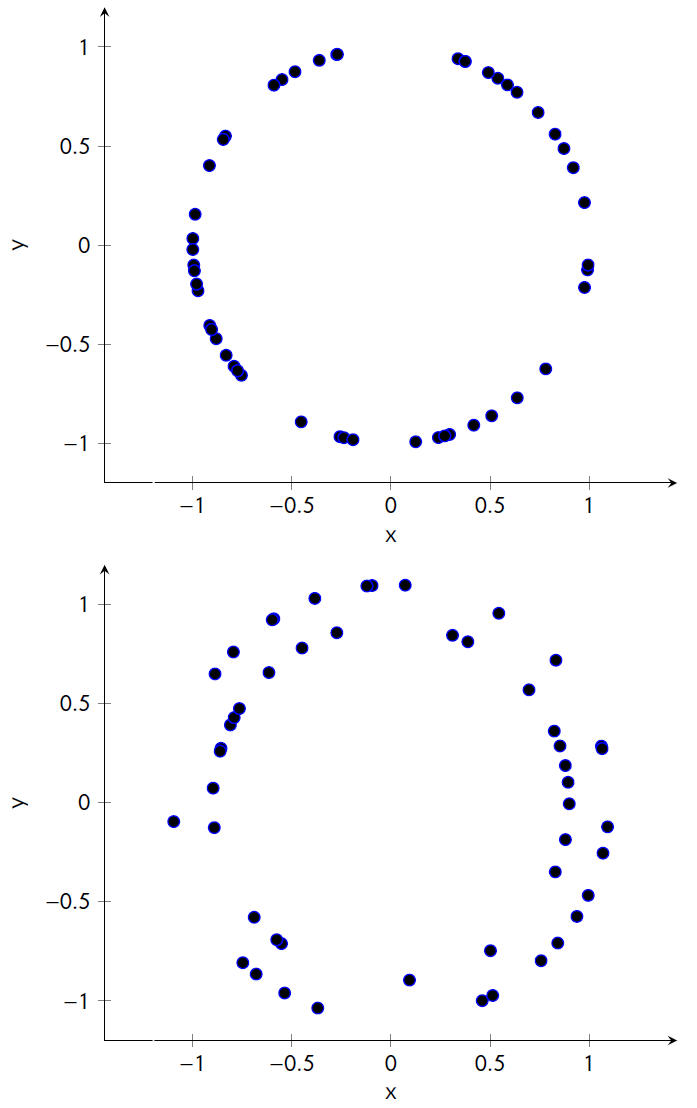}
			\caption{Scatterplots}\label{subfig:scatter}
		\end{subfigure}
		\begin{subfigure}{0.23\textwidth}
			\centering
			\includegraphics[height=1.22\linewidth]{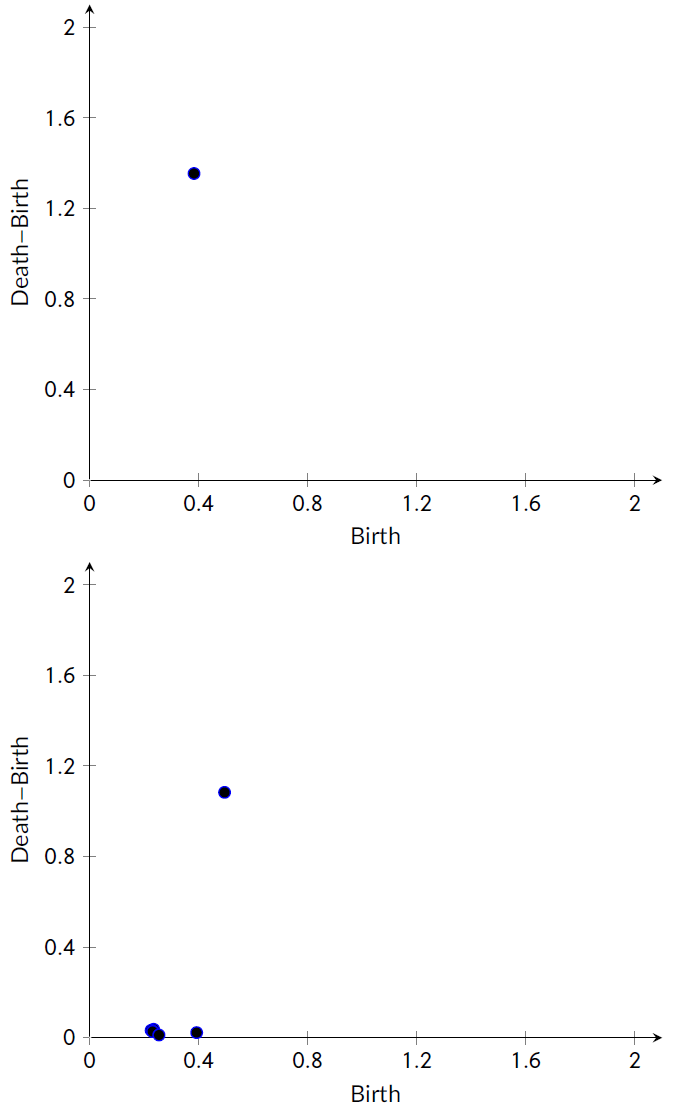}
			\caption{PDs}\label{subfig:PD}
		\end{subfigure}
	    \begin{subfigure}{0.23\textwidth}
			\centering
			\includegraphics[height=1.22\linewidth]{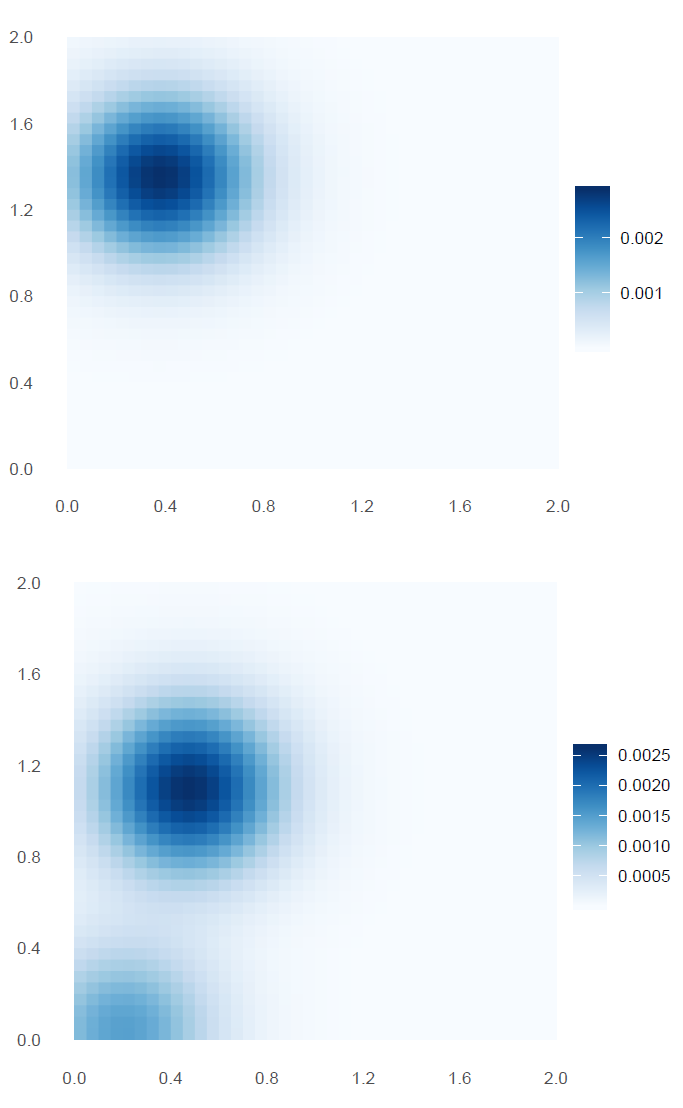}
			\caption{PIs}
			\label{subfig:PI}
     	\end{subfigure}
		\caption{(a) Two shapes, (b) scatterplots of 50 randomly sampled points, (c) transformed dimension-one persistence diagrams, and (d) persistence images. We use the Gaussian smoothing function with the arctangent weight for the persistence images.}
		\label{fig:sim}
	\end{figure}

\begin{figure*}[!t]
		\begin{subfigure}{0.32\textwidth}
			\centering
			\includegraphics[width=1\linewidth]{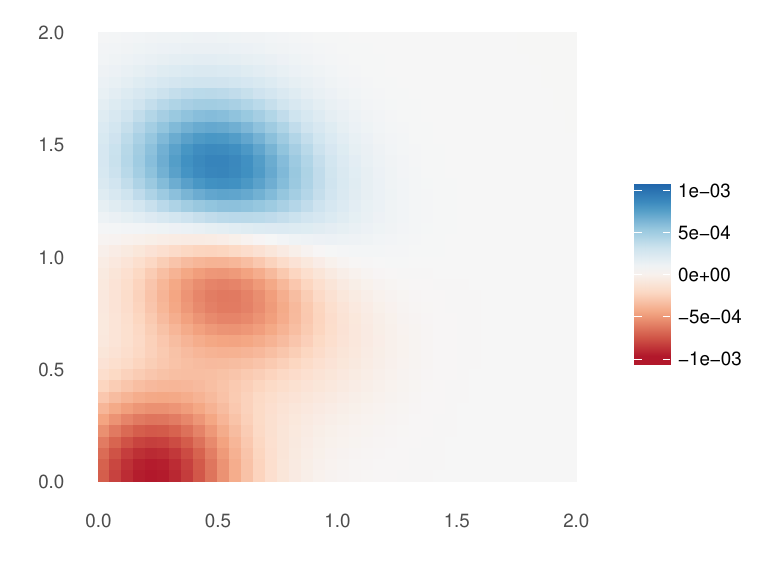}
			\caption{Average difference}
			\label{subfig:ttestdiff}
		\end{subfigure}
		\begin{subfigure}{0.32\textwidth}
			\centering
			\includegraphics[width=1\linewidth]{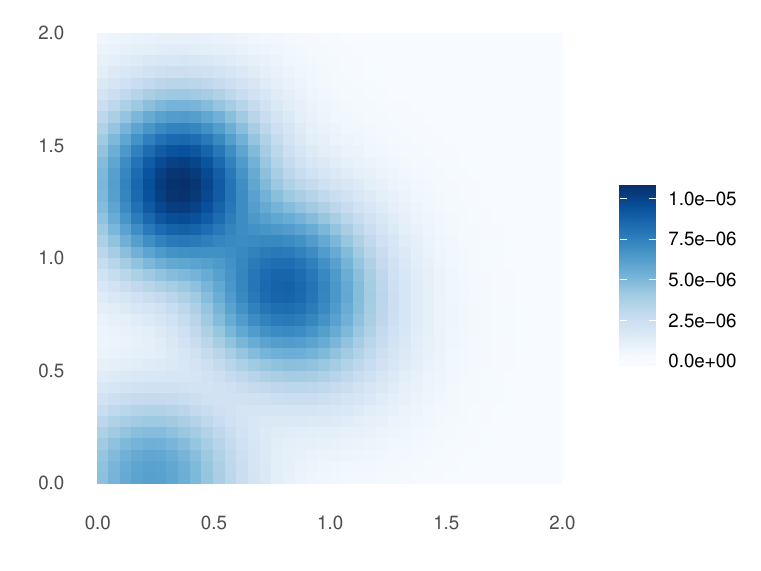}
			\caption{Standard errors}
			\label{subfig:ttestse}
		\end{subfigure}
		\begin{subfigure}{0.32\textwidth}
			\centering
			\includegraphics[width=0.9\linewidth]{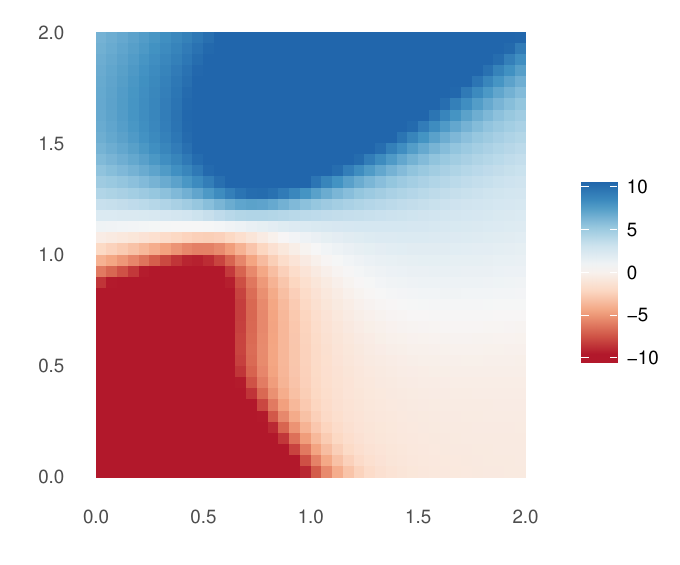}
			\caption{$t$-test statistics}
			\label{subfig:ttestt}
		\end{subfigure}
		\caption{Average differences of one-dimensional persistence images, standard errors, and $t$-test statistics. The $t$-test statistics less than -10 and greater than 10 are plotted as -10 and 10, respectively.}
		\label{fig:sim.res}
	\end{figure*}

	To illustrate these issues, we reproduce a simulation study considered in \cite{Robinson2017}. We randomly sample 100 sets of 50 points from each of two different shapes: one circle of radius 1 (shape 1) and two circles of radius 0.9 and 1.1 (shape 2), as shown in Figure~\ref{fig:sim}. 
	We compute persistent homology for 200 sets of point cloud data using the Rips complex.
	The difference between the two spaces is reflected in the dimension-one persistence diagrams, given in the second from the right panel of Figure~\ref{fig:sim}. The dimension-one interval for shape 1 ($ \text{Death}- \text{Birth} \approx1.3$) is longer than the longest dimension-one interval for shape 2 ($ \text{Death}- \text{Birth} \approx1.1$). Also, shape 2 has several very short-lived dimension-one intervals.
	The 200 dimension-one persistence diagrams are converted to 200 persistence images of $40 \times 40$ pixels using the Gaussian smoothing function with $h=0.075$ and the arctangent weight.
	We conduct the two-sample pooled variance $t$-test at every pixel to examine the differences between the two spaces. 
	Figure~\ref{fig:sim.res} presents the average difference of the persistence images, standard errors, and two-sample $t$-test statistics.

	In the naive hypothesis testing, some pixels irrelevant to inference are used.
	Including unnecessary pixels could reduce the detection power of the test.
	Even though some pixels have small mean differences, they can have large test statistics.
	For example, some pixels on the top in Figure~\ref{subfig:ttestt} have large test statistics but they have small mean differences in Figure~\ref{subfig:ttestdiff}. This is because their sample standard deviations are relatively small in Figure~\ref{subfig:ttestse}.
	As a result, it is difficult to differentiate whether the large test statistics are due to the mean differences or the small sample standard errors.
	
	Also, when we test multiple hypotheses simultaneously, the multiple testing problem arises. Failure to adjust for multiple testing can lead to inaccurate inference. Multiple testing adjustments based on the family-wise error rate \citep{Hochberg1987} and the FDR \citep{Benjamini1995} aim to control the rate of false positives while maintaining statistical power.

	In the next section, we suggest a hypothesis testing approach for persistent images to handle irrelevant and misleading pixels and adjust for conducting multiple tests.

	\section{Two-stage Hypothesis Test Procedures}
	\label{sec:ts}
	 The suggested hypothesis test approach consists of two stages: filtering and testing.
	 We use two independent statistics (filter and test statistics) in the two stages. Algorithm~\ref{alg:vec} summarizes the suggested testing procedure.

	 \begin{algorithm}[!t]
		\caption{Two-stage hypothesis test for a persistence image}\label{alg:vec}
		\begin{algorithmic}[1]
			\Require {Array of $n$ persistence images $V_{\{m^2 \times n\}}$, given label $G_{\text{given}}=\{I,J\}$, threshold $C$}
			\Ensure{P-values $Z$}
			\State {Create vectors $v_x,v_y$ of size $m^2$ that correspond to $x$ and $y$ locations of $V$}
			\State {$V \leftarrow V [ v_x \geq v_y , : ]$ }
			\State {Create an empty vector $T$ of size $ \frac{m(m+1)}{2}$}
			\For {$i = 1 \to \frac{m(m+1)}{2}$}
			\State {$T[i]\leftarrow$ filter statistic of $i$th pixel of $V$}
			\EndFor
			\State $t_C \leftarrow C^{th}$ percentile of $T$
			\State $V\leftarrow V[V>t_C,:]$
			\State {Create an empty vector $Z$ of size $nrow(V)$}
			\For {$j=1 \to nrow(V)$}
			\State $v_1\leftarrow V[j,I]$
			\State $v_2\leftarrow V[j,J]$
			\State {Conduct a hypothesis test using $v_1$ and $v_j$ and store the p-value in $Z[j]$}
			\EndFor
			\State {Apply multiple testing adjustment procedure to $Z$}
		\end{algorithmic}
	\end{algorithm}

	\subsection{Stage I: Filtering} The idea of filtering has been proposed to increase power for high-dimensional data \citep{McClintick2006,Hackstadt2009,Mieth2016}. 
	When the variables are filtered, the number of variables being tested is reduced. As a result, filtering could potentially lead to an increase in the number of discoveries after multiple testing corrections.

	We add the pre-filtering step of removing unnecessary pixels in the upper-right triangle of the persistence image. This part of the persistence image corresponds to an empty area in the persistence diagram. For example in Figure~\ref{fig:PI} the upper-right triangle region in the transformed persistence diagram in the middle panel  corresponds to the empty region of $\text{Death}>1$ in the persistence diagram in the left panel. Therefore, the upper-right triangle region in the transformed persistence diagram should not contribute to inference. After applying the pre-filtering step to the persistence image of $m \times m=m^2$ pixels, the number of remaining pixels is reduced to $\frac{m(m+1)}{2}$.

	\cite{Bourgon2010} show that false positive rates are not maintained for the two-stage procedure if inappropriate filter statistics are used. 
	More specifically, filter statistics need to be independent of the test statistic because the null distribution in the second stage is a conditional distribution given the filter statistic. For example, the overall sample mean, overall standard deviation, and sum of squared values are suggested as filter statistics for the $t$-test statistic \citep{Bourgon2010,Guo2017}.
	In our study, we use the overall standard deviations $s^i=\sqrt{ \frac{\sum\limits_{j=1}^2 \sum\limits_{k=1}^{n_j}\left(x_{(j,k)}^{i}- \bar{x}^i \right)^2}{n_1+n_2-1}}$, where $\bar{x}^i=\frac{1}{n_1+n_2}\sum\limits_{j=1}^2 \sum\limits_{k=1}^{n_j} x_{(j,k)}^{i}$, as filter statistics for each pixel $i\in\left\{1,\ldots,\frac{m(m+1)}{2} \right\}$. 
	For a given threshold $C$ where $0\leq C \leq 100$, pixels whose filter statistics are less than the $C^{th}$ percentile are removed.

	\subsection{Stage II: Testing}	
	In the second stage, we conduct hypothesis tests on the remaining pixels and adjust for multiplicity. 
	Though we use the pooled-variance $t$-test in our analysis, any other standard approach can be used, such as unpooled $t$-test, nonparametric two-sample tests, ANOVA, and nonparametric ANOVA \citep{Bourgon2010}. No additional difficulties are introduced by the transition from two to multiple groups as long as a suitable filter statistic is used. For example, one may test whether at least one of the structures of the four rocks in Figure~\ref{fig:rocks} are different or not using ANOVA. We also use ANOVA to compare three groups of material images in Section~\ref{subsec:2drock}.

	The adjusted multiple hypothesis results could be affected by the dependency between persistence image pixels. Also, filtering can change the dependence structure of the pixels that pass the filter \citep{Bourgon2010}. Widely-used multiple testing adjustment methods assume independence between tests \citep{Benjamini1995,Storey2002}, whereas some methods control the error rates under various types of dependence \citep{Benjamini2001,Kim2008,Fan2012,Stevens2017}. 
	In our study, we adjust the p-values using the BH method \citep{Benjamini1995}, the BY method \citep{Benjamini2001}, and the q-value method of \cite{Storey2002}. For more detailed explanation of these multiple testing adjustment methods, see Section~S3 of the supplementary material. 
	
	The effects of hypothesis testing settings such as the choice of the filtering threshold $C$ and the multiple testing adjustment methods are studied in Section~\ref{subsec:parameter}.

	\section{Simulation Study}
	\label{sec:sim}
	
	\subsection{Method Comparison}
	\label{subsec:comparsion}

	We compare the proposed two-stage method using persistence images with three persistent homology based hypothesis testing methods: 1) permutation tests using persistence diagrams \citep{Robinson2017} (PD), 2) permutation tests using persistence landscape \citep{Bubenik2015} (PL), and 3) kernel two-sample tests applied to the persistent weighted Gaussian kernel (PWGK) \citep{Kusano2019} (Kernel). 
	For the two-stage hypothesis tests, we use persistence images of 40 by 40 pixels with Gaussian smoothing of $h=0.5$ and constant weight. The overall sample standard deviations are used as the filter statistics and the filtering threshold is set at $C=80\%$. The p-values are adjusted by the BH procedure. 
	In our study, PWGK for the Kernel method is generated by the same smoothing kernel and weight. 
	Also, the null distribution for the kernel two-sample test is approximated using $1,000$ bootstrap samples on the aggregated data.
	For PD and PL, 1,000 permutations ($N_P=1,000$) are used. For PD, pairwise distances between persistence diagrams are measured by the 1-Wasserstein distance.

	Two criteria are used to evaluate the performances of the tests: false positive rate and power.
	A good hypothesis test will achieve high power while maintaining low false positive rate. 
	We use the same simulation setting used in \cite{Robinson2017}, the point cloud data sampled from the two shapes of Figure~\ref{fig:sim}.

	First, for simulations to examine the false positive rate, we randomly draw 20 point clouds of 50 points on shape 2 (two different-sized circles) and add noise that follows $N(0,\sigma^2)$.
	These 20 point clouds are randomly assigned to two groups of the same size, 10 point clouds to each group, respectively. 
	We repeat this procedure 500 times for a given $\sigma=0.05$, 0.1, 0.15, and 0.2.
	
	Second, to evaluate the power of the hypothesis tests, we randomly draw 10 point clouds of 50 points on shape 1 (one circle) and add Gaussian noise of $N(0,\sigma^2)$.
	Then, we apply the same sampling procedure for shape 2 (two different-sized circles). 
	These 20 point clouds, ten from each shape, are used to conduct one hypothesis test. We repeat this procedure 500 times for $\sigma=0.05$, 0.1, 0.15, and 0.2.
	
	We construct the Rips complex separately for each of the $20\times500\times4\times2=80,000$ point clouds and compute persistent homology. 
	For the hypothesis tests, the dimension-one persistence diagrams are used.

	\begin{figure*}[!ht]
		\centering
		\begin{subfigure}{0.32\textwidth}
			\centering
			\includegraphics[width=1\linewidth]{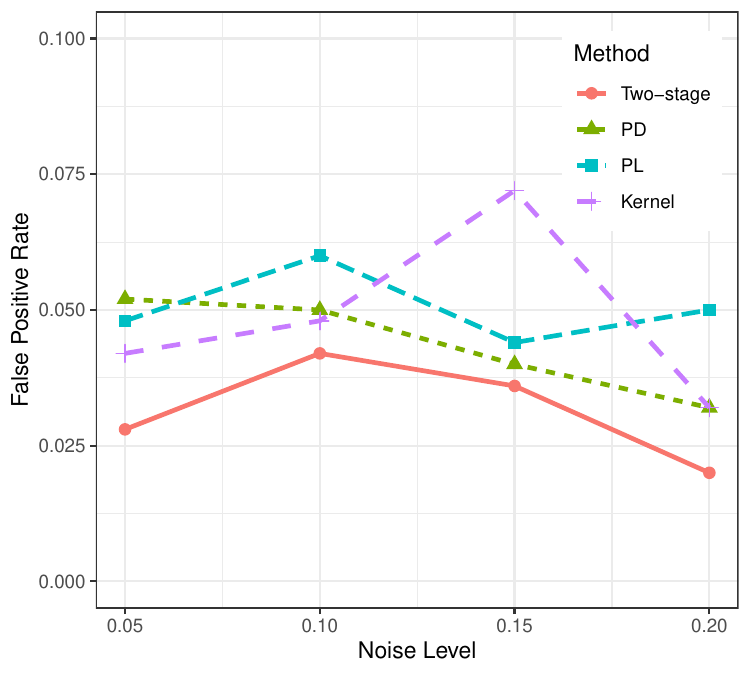}
			\caption{False positive rate}
			\label{subfig:fpr}
		\end{subfigure}
		\begin{subfigure}{0.32\textwidth}
			\centering
	\includegraphics[width=1\linewidth]{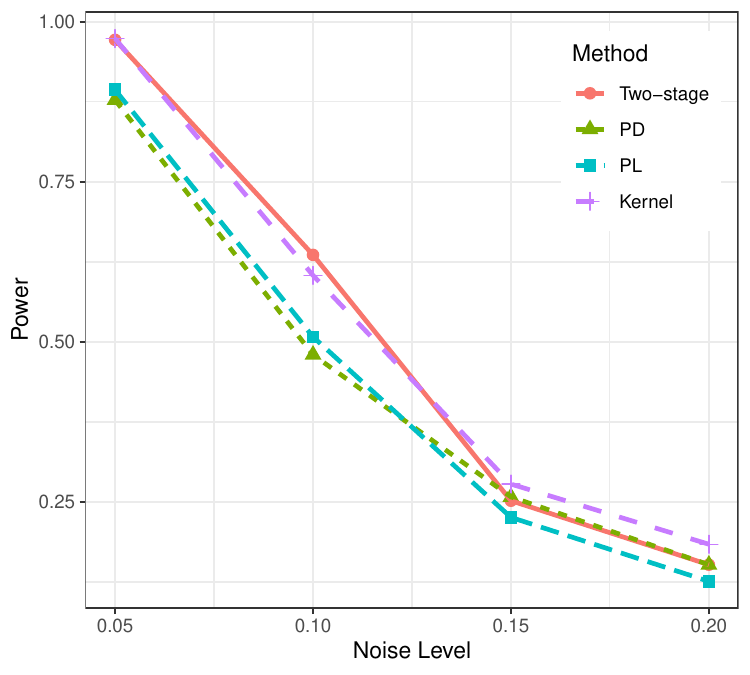}
			\caption{Power}
			\label{subfig:power}
		\end{subfigure}
		\begin{subfigure}{0.32\textwidth}
			\centering
	\includegraphics[width=1\linewidth]{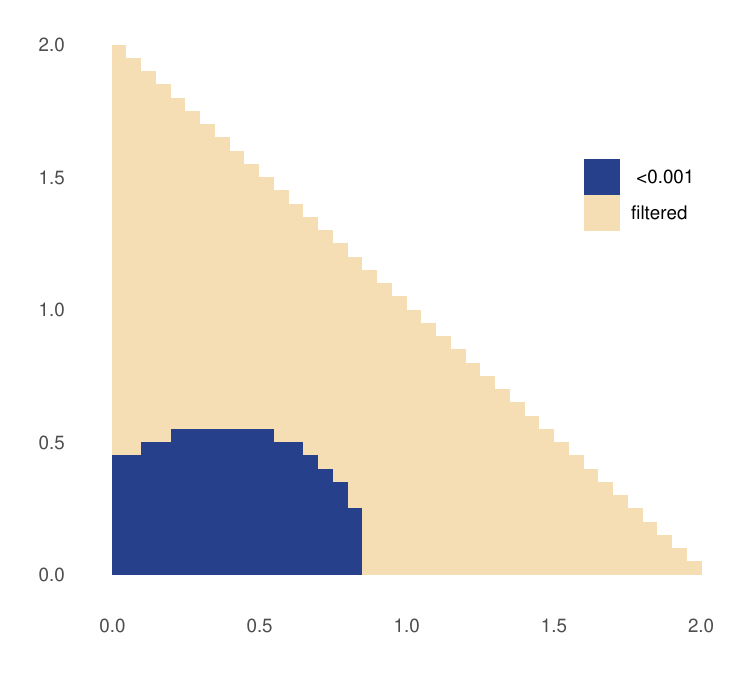}
			\caption{p-values of two-stage test}
			\label{subfig:qval80}
		\end{subfigure}
		\caption{Hypothesis testing results of the four methods at significance level $\alpha=0.05$: (a) false positive rates; (b) powers; (c) p-values adjusted by the BH method plotted on the dimension-one persistence image with $\sigma=0.05$, constant weight, and $C=80$\%. PD is the permutation test of \cite{Robinson2017}, PL is the permutation test of \cite{Bubenik2015}, and Kernel is the kernel two-sample test of \cite{Kusano2019}.}
		\label{fig:method}
	\end{figure*}
	
	Figures~\ref{subfig:fpr} and~\ref{subfig:power} show false positive rates and powers of the four testing methods (Two-stage, PD, PL, and Kernel). 
	The simulation results suggest that the proposed two-stage method achieves high power while maintaining the lowest false positive rates.

	In addition, the proposed two-stage method can inform which topological features contribute the most to the differences. For example, Figure~\ref{subfig:qval80} shows the adjusted p-values of filtered pixels of the dimension-one persistence image of one simulated data set for two shapes when $\sigma=0.05$, constant weight, and $C=80$\% are used. The pixels that have small p-values are located at the lower-left corner of Figure~\ref{subfig:qval80}, suggesting that the differences between the two groups of point clouds can be identified by the number of small-sized loops. 
	For example, the sampled points in the bottom panel of Figure~\ref{subfig:scatter} have a narrow space between the two circles and they generate the small-sized loops. On the other hand, the point cloud sampled from the one circle in the top panel of Figure~\ref{subfig:scatter} does not have such space and generates a single large loop.

	\subsection{Effects of Two-stage Hypothesis Testing Settings}
	\label{subsec:parameter}
	
	The proposed two-stage hypothesis test depends on multiple parameters and settings. We examine the effects on testing results of four conditions: 1) filtering threshold, 2) weight, 3) multiple testing adjustment method, and 4) resolution of the persistence image. The effect of persistence image resolution is presented in Figure~S7 in the supplementary material. We use the same point cloud data sets used in Section~\ref{subsec:comparsion}. 
	
	\subsubsection{Effect of Filtering Thresholds}
	\label{subsubsec:filtering}

\begin{figure*}[!t]
		\centering
		\begin{subfigure}{1\textwidth}
		\centering
        \includegraphics[width=0.36\linewidth]{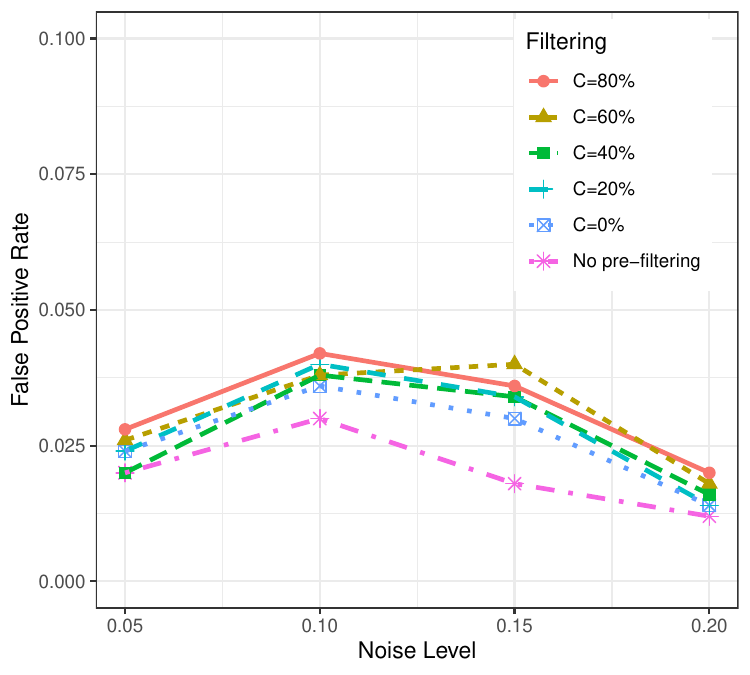}
        \includegraphics[width=0.36\linewidth]{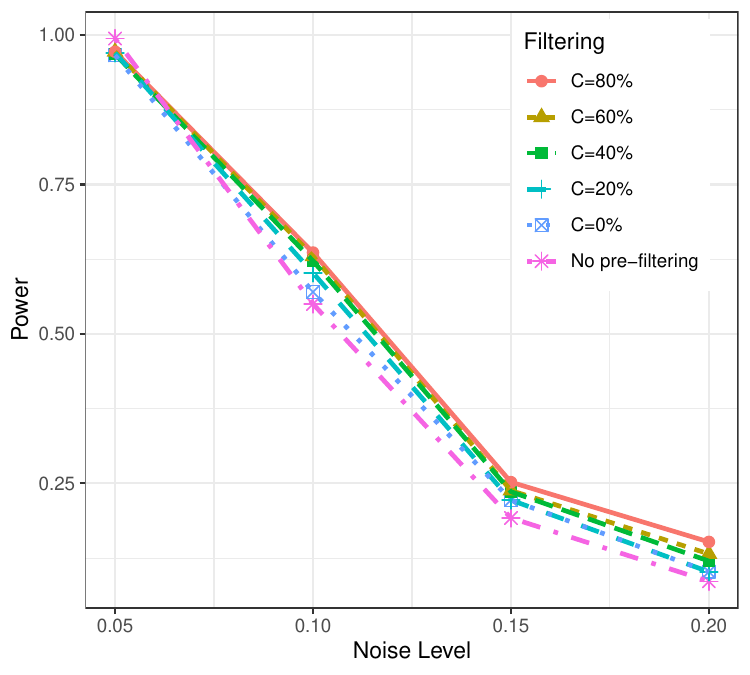}
        \caption{False positive rate (left) and power (right) at significance level $\alpha=0.05$ of six filtering conditions.}
        \label{fig:filtering}
        \end{subfigure}
        \begin{subfigure}{1\textwidth}
        \centering
        \includegraphics[width=0.36\linewidth]{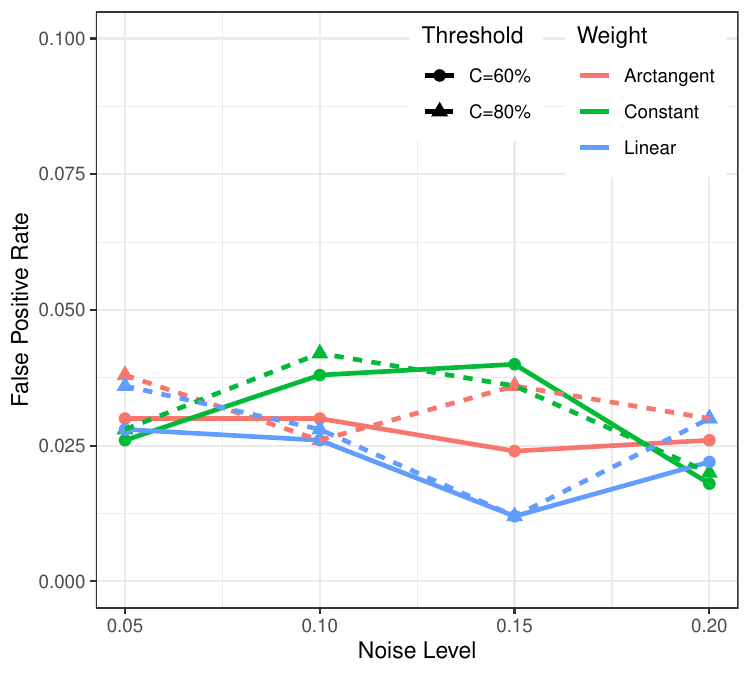}
        \includegraphics[width=0.36\linewidth]{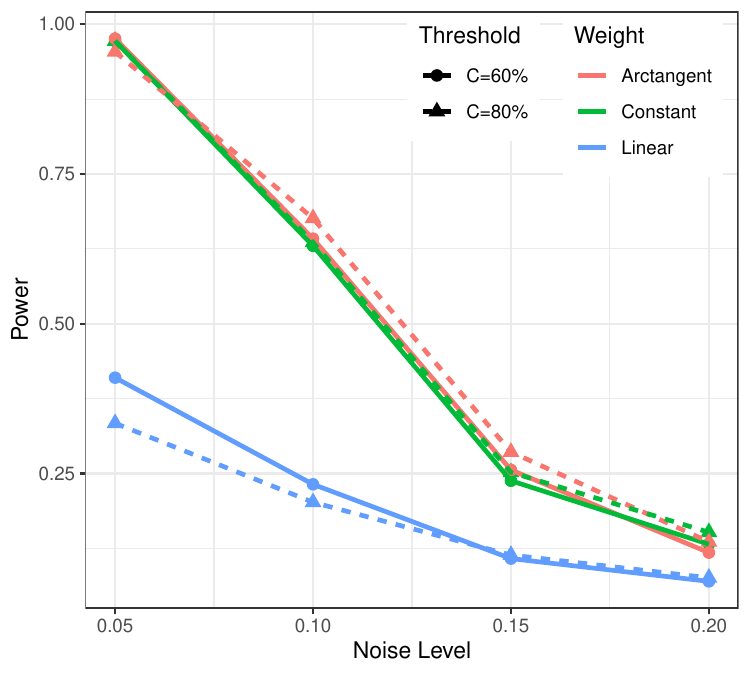}
        \caption{False positive rate (left) and power (right) at significance level $\alpha=0.05$ of three weights.}
        \label{fig:weight}
        \end{subfigure}
        \begin{subfigure}{1\textwidth}
        \centering
        \includegraphics[width=0.36\linewidth]{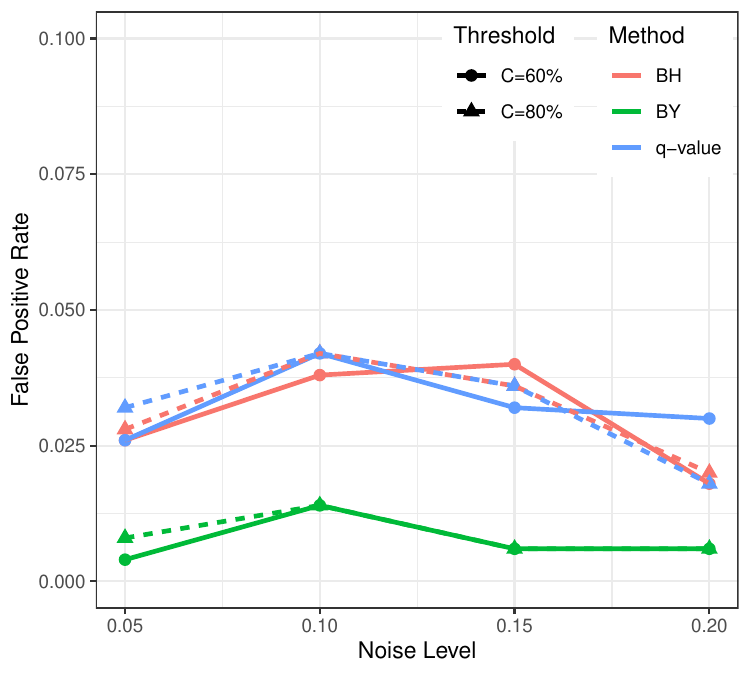}
        \includegraphics[width=0.36\linewidth]{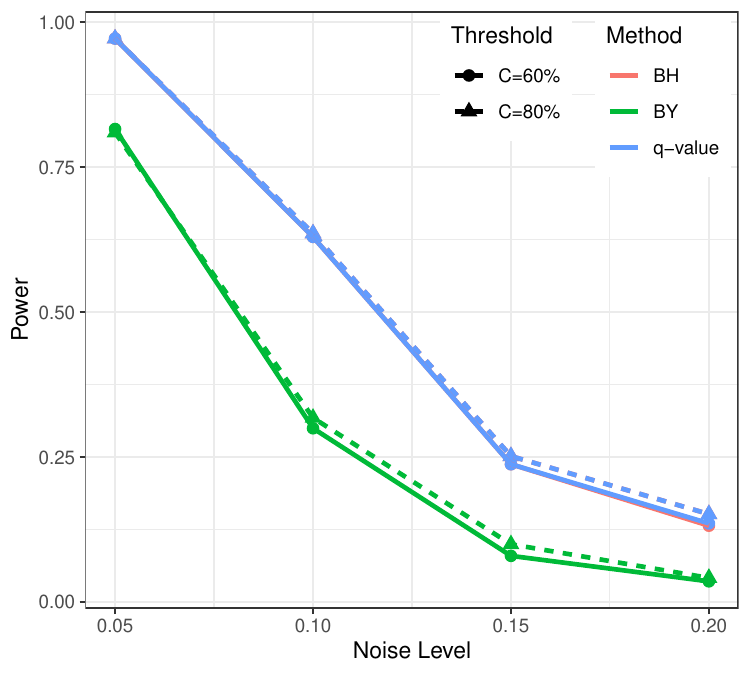}
        \caption{False positive rate (left) and power (right) at significance level $\alpha=0.05$ of three multiple testing adjustment methods.}
        \label{fig:control}
        \end{subfigure}
        \caption{Simulation results in various two-stage hypothesis testing settings: (a) filtering threshold; (b) weight; and (c) multiple testing adjustment method.}
	\end{figure*}

	We explore six filtering conditions: no pre-filtering (using all $40 \times 40=1,600$ pixels of persistence images) and five filtering thresholds $C=0$\% (using $40\times41/2 =820$ pixels), 20\% (using $40\times41/2\times(1-0.2) =656$ pixels), 40\%, 60\% and 80\%. Figure~\ref{fig:filtering} shows the false positive rate and power for each of the six filtering conditions. 
	
	The simulation results indicate that in general, power increases with the amount of filtering. 
	When no pre-filtering is used, the procedure has the lowest power. On the other hand, the filtering procedure increases the power of the test.
	Also, using higher filtering thresholds tends to increase the false positive rates. However, they are controlled less than the nominal level 0.05 for all filtering conditions.
	Note that using higher filtering thresholds may not always lead to higher powers as shown in Figure~\ref{fig:filtering}. For example, if a majority of important pixels are filtered out using a high filtering threshold, the detection power could decrease. 
	
	Figure~S8 in the supplementary Material shows the BH-adjusted p-values of the persistence images of one simulated dataset with $\sigma=0.05$ for the six filtering settings, no pre-filtering and $C=0$\%, 20\%, 40\%, 60\% and 80\%.  
	
	\subsubsection{Effect of Weights}
	We compare three weights, the constant, arctangent, and linear weights, used in the vectorization procedure.
	The constant weight gives the same weight to features regardless of their persistence. 
	On the other hand, the linear weight tends to assign higher weights to longer-persistent features (i.e., points that are far from the 45-degree line in the persistence diagrams). The arctangent weight is somewhere between the constant and linear weights.

	The weights can play an important role in determining detection power.
	Figure~\ref{fig:weight} compares power of two-stage hypothesis tests with filtering thresholds $C=60$\% and 80\% under the three weights.
	For this example, the constant and arctangent weights yield higher power than the linear weight. 
	In our simulated datasets, short-lived one-dimensional topological features (i.e., features close to the 45-degree line in the persistence diagrams) play an important role in differentiating the two groups.
	Because the linear weight assigns larger weights to persisting features, it does not capture such differences compared to the constant and arctangent weights.
	
	The results imply that the proposed two-stage test provides flexible options to compare the differences between collections of persistence diagrams. 
	The permutation tests of \cite{Robinson2017} and \cite{Bubenik2015} can be viewed as providing limited flexibility because they cannot assign different weights to the topological features.

	\subsubsection{Effect of Multiple Testing Adjustments}

    We compare three multiple testing adjustment methods: the BH method that assumes independence between tests \citep{Benjamini1995}, the BY method that assumes positive regression dependence \citep{Benjamini2001}, and the q-value method of \cite{Storey2002} that is known to work well under more general dependence structure than positive regression dependence \citep{storey2004strong}. For the q-value method, when the proportion of true null hypotheses is not estimated, the BH method is used. Figure~\ref{fig:control} reports simulation results of the three multiple testing adjustment methods. 
    
    In this simulation study, the BH method performs the best in the sense that false positive rate is well controlled while maintaining high power. The BY method records the lowest false positive rate but has the lowest power. 
    The q-value method yields almost the same results to the BH method. This is because the proportion of the true null hypotheses is not estimated in most cases, so the BH method is used instead.

	\subsection{2D Binary Material Image Simulation}
	\label{subsec:2drock}
	
\begin{figure}[!ht]
		\centering
		\begin{subfigure}{0.25\textwidth}
			\centering
			\includegraphics[width=1\linewidth]{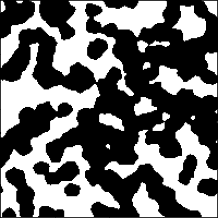}
		\end{subfigure}
		\begin{subfigure}{0.25\textwidth}
			\centering
			\includegraphics[width=1\linewidth]{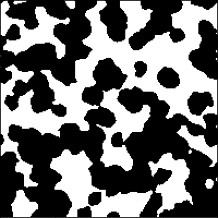}
		\end{subfigure}
		\begin{subfigure}{0.25\textwidth}
			\centering
			\includegraphics[width=1\linewidth]{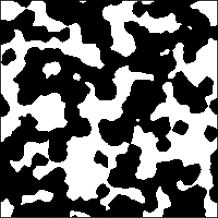}
		\end{subfigure}
		\caption{Examples of 2D pseudo-material images with parameters $(M=180,S=80)$ (left), $(M=190,S=75)$ (center), and $(M=200,S=70)$ (right).}
		\label{fig:sim.rock}
	\end{figure}

	In this section, we simulate pseudo-material data, binary images with pores and grains.
	The 2D random binary images are generated by Algorithm 1 of \cite{Obayashi2018}. First, $M$ seed points are taken from a uniform distribution over the image. From each seed point, $S$ dispersion points are randomly generated from $N(0,\sigma_1^2)$. A Gaussian filter with standard deviation $\sigma_2$ is applied and binarized using a threshold $t$. 
	Three sets of parameters are used to generate images: $(M=180,S=80)$, $(M=190,S=75)$, and $(M=200,S=70)$.
	For all settings, we set dispersion scales $\sigma_1=\sigma_2=4$ and threshold $t=0.7$ and generate images of size $200$ by 200 pixels. Figure~\ref{fig:sim.rock} shows examples of simulated binary images from each parameter set. It is not easy to tell that the simulated images are generated by different parameters by visual inspection alone. 
	
	We consider four scenarios to compare the pseudo-material images: scenario 1 examines two groups of $(M=180,S=80)$ images, scenario 2 compares $(M=180,S=80)$ and $(M=190,S=75)$ groups, scenario 3 tests $(M=180,S=80)$ and $(M=200,S=70)$ groups, and scenario 4 tests whether at least one of the three groups, $(M=180,S=80)$, $(M=190,S=75)$, and $(M=200,S=70)$, differs from the others.
	All groups in each scenario include 50 images of size 200 by 200. 
	The binary images are converted by the signed distance transform and cubical complexes are constructed according to the signed distance values. 
	The topological features obtained by this transformation can reveal the size, shape, and connectivity of materials  \citep{Robins2016,Obayashi2018}.
	A more detailed description and examples of the topological features of 2D binary material images are given in Section~S2.3 of the supplementary material.
	Persistent homology is computed using the GUDHI library \citep{gudhi} with the SEDT values as the filtration.

	Four hypothesis testing methods (two-stage, PD, PL, and Kernel) are applied to the computed topological features of the pseudo-material images. For the two-stage hypothesis tests, we use the persistence images of size 40 by 40 converted using three weights (linear, arctangent, and constant) with Gaussian smoothing of $h=3$, overall standard deviation filter statistic, filtering threshold $C=60\%$, and the BH method. PD and PL use $N_P=100$ and $N_P=1,000$ repetitions, respectively. For Kernel, the arctangent weight, the Gaussian kernel of $h=3$, and 1,000 bootstrap samples are used.
	
		\begin{table*}[!ht]
		\caption{p-values of four hypothesis testing methods of 2D binary material images. The minimum p-values are reported for the two-stage tests. 
		}
		\begin{tabular*}{\textwidth}{@{\extracolsep{\fill}}llcccccccc@{\extracolsep{\fill}}}
				\hline
				&        & \multicolumn{2}{c}{Scenario 1} & \multicolumn{2}{c}{Scenario 2} & \multicolumn{2}{c}{Scenario 3} & \multicolumn{2}{c}{Scenario 4} \\ \cline{3-10} 
				&        & Dim 0              & Dim 1             & Dim 0                 & Dim 1                 & Dim 0                 & Dim 1         & Dim 0                 & Dim 1            \\ \hline
				Two-stage & Linear & 0.547              & 0.993             & 0.997                 & 0.127                 & 0.005                 & 0.032    & 0.005 &  0.053           \\ 
				\multicolumn{1}{l}{}               & Arctangent   & 0.516              & 0.999             & 0.998                 & 0.133                 & $<$0.001                 & 0.021   & $<$0.001 &  0.030           \\  
				\multicolumn{1}{l}{}               & Constant     & 0.549              & 0.992             & 0.999                 & 0.139                 & $<$0.001                 & 0.021  &  $<$0.001 &  0.026             \\ 
				\multicolumn{2}{l}{PD}        &   0.400  &     0.990      &    0.900                   &   0.070                    &         0.000              &   0.010    & - & -                 \\ 
				\multicolumn{2}{l}{PL}        &   0.319  &   0.932        &    0.978                   &   0.076                    &         0.007              &   0.018   & - & -                 \\ 
				\multicolumn{2}{l}{Kernel}        &   0.000 &   0.000        &    0.000                   &   0.000                    &         1.000             &   0.000   & - & -                 \\ \hline
			\end{tabular*}
	\label{tb:rocksim}
	\end{table*}

	The hypothesis test results show that the proposed method can differentiate the simulated binary images based on their topological features. 
	Table~\ref{tb:rocksim} shows p-values from hypothesis tests. Here, the minimum p-value is selected as a representative measure for the two-stage tests to compare performances with the other three methods. 
	The two-stage test, PD, and PL show similar testing results: scenario 1 and 2 yield larger p-values whereas scenario 3 yields small p-values. 
	We note that the two-stage test provides more information with multiple p-values corresponding to the filtered pixels. 

	On the other hand, Kernel shows the most extreme testing results; the p-values are either 0 or 1 for all scenarios. 
	For scenario 4, only two-stage tests are used because the other tests are developed to compare two groups. We use ANOVA and its p-values are similar to those of scenario 3.
	
	\begin{figure}[!t]
		\centering
		\begin{subfigure}{0.32\textwidth}
			\centering
			\includegraphics[width=1\linewidth]{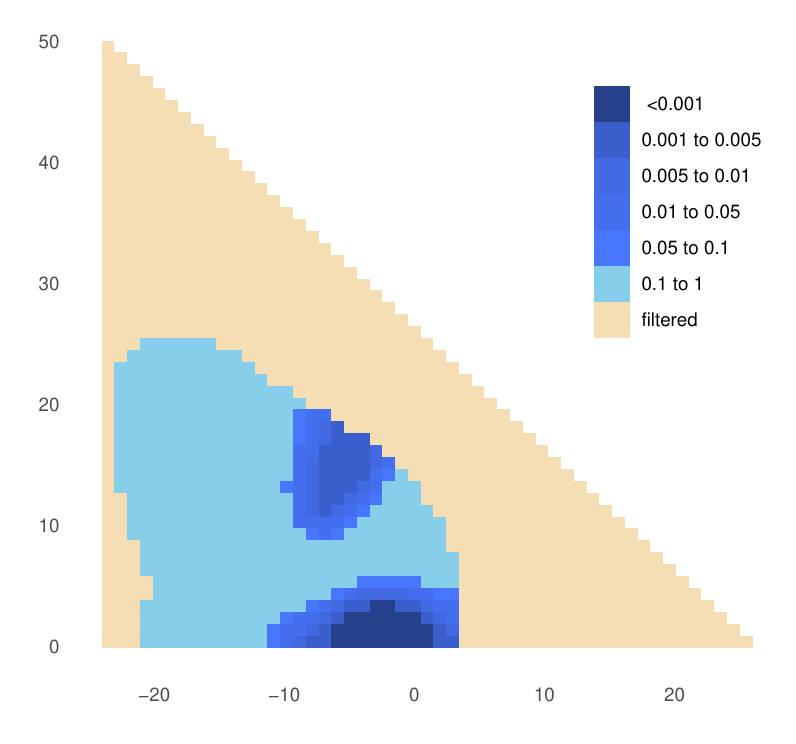}
			\caption{Dimension-zero}
			\label{subfig:rocksimdim0}
		\end{subfigure}
		\begin{subfigure}{0.32\textwidth}
			\centering
			\includegraphics[width=1\linewidth]{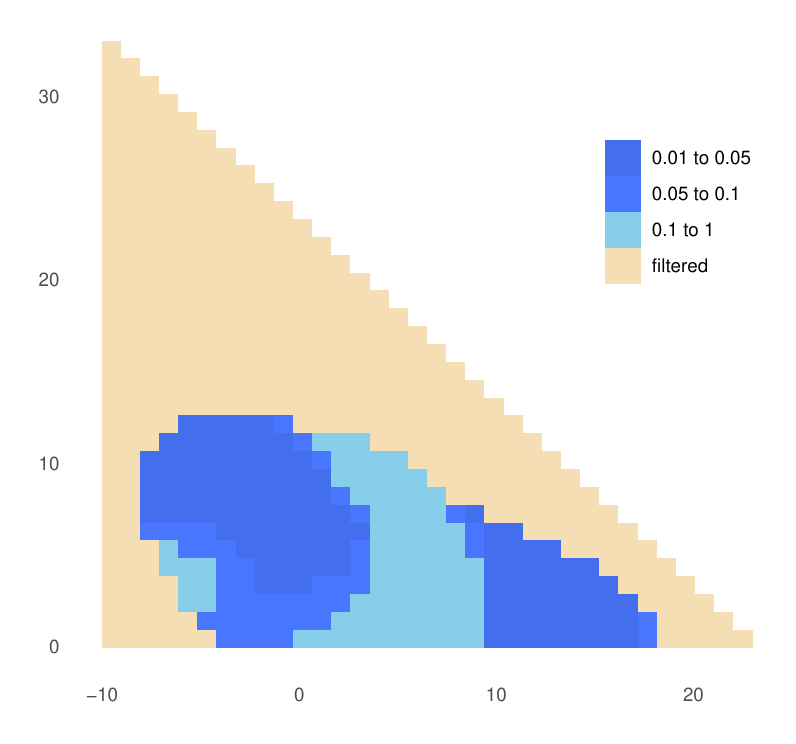}
			\caption{Dimension-one}
			\label{subfig:rocksimdim1}
		\end{subfigure}
		\caption{p-values of the two-stage hypothesis test of scenario 3, between pseudo-rock images of $(M=180,S=80)$ and $(M=200,S=70)$ groups.}
		\label{fig:rocksim}
	\end{figure}

	Unlike the other methods, the two-stage approach shows which topological features play an important role in hypothesis tests. Figure~\ref{fig:rocksim} shows the BH-adjusted p-values of two-stage hypothesis test of scenario 3 with the arctangent weight and $C=$60\%. 
	For dimension-zero, two regions have small p-values around $(-5,0)$ and $(-5,15)$ in Figure~\ref{subfig:rocksimdim0}. The areas at $(-5,0)$ and $(-5,15)$ correspond to the connected grains and disconnected grains, respectively. See panel (e) of Figure~S6(a) in the supplementary material for an example of the corresponding dimension-zero feature. 
	For dimension-one, there are two areas with small p-values: 1) around $(-5,10)$ and 2) around $(15,5)$ in Figure~\ref{subfig:rocksimdim1}. The areas at $(-5,10)$ and $(15,5)$ correspond to the pores and the broken-ring shaped grains, respectively. See panels (a) and (b) of Figure~S6(b) in the supplementary material for examples of the corresponding dimension-one features. 
	As a result, the proposed two-stage test provides information on how the two groups differ, which are not easily identifiable in the binary material images in Figure~\ref{fig:sim.rock}. The other testing methods do not provide this level of detail. This is because they are based on permutations, so output is a single p-value.

	\subsection{Population of Tribolium Beetle}
	\label{subsec:beetle}
	
	In this section, we apply the proposed method to the simulated beetle population data.
	Tribolium, also known as a flour beetle, is a pest that infests stored food products \citep{mason2012biology}. Tribolium is considered to have great economic importance because it is globally spread and resistant to several pesticides \citep{verheggen2007electrophysiological}. The population dynamics of Tribolium have been studied using experimental and mathematical models \citep{costantino2005nonlinear}.

	Tribolium has four life stages: 1) egg, 2) larva, 3) pupa, and 4) adult. In each of the larva and pupa stages, it takes about two weeks for Tribolium to get to the next stage. The first three to four days of adults are immature and nonproductive, so they are called the callow adult. 
	Tribolium has a characteristic of cannibalism, where adults eat pupae and unhatched eggs under overpopulation conditions.
	We use the Tribolium population growth model proposed in \cite{costantino1995experimentally}.
	The population model consists of the following equations:
	\begin{eqnarray*}
	    L_{t+1}&=&bA_t\exp(-c_{ea}A_t-c_{el}L_t+E_{1t}) \\
	    P_{t+1} &=& L_t(1-\mu_l)\exp(E_{2t})\\
	    A_{t+1}&=&\left[ P_t\exp(-c_{pa}A_t)+A_t(1-\mu_a) \right]\exp(E_{3t}),
	\end{eqnarray*}
	where $L_t$ is the number of feeding larvae, $P_t$ is the number of non-feeding larvae, pupae and callow adults, and $A_t$ is the number of mature adults at time $t$. The population numbers are recorded every two weeks to equate to the feeding larval maturation cycle. The parameter $b>0$ is the number of larval recruits per adult per unit of time when there is no cannibalism, $\mu_l$ and $\mu_a$ are the proportions of the larvae and adults that die from non-cannibalism causes, $\exp(-c_{pa}A_t)$ is the survival probability of pupae when there are $A_t$ adults, and $\exp(-c_{ea}A_t)$ and $\exp(-c_{el}L_t)$ are the probabilities that an egg is not eaten when there are $A_t$ adults and $L_t$ larvae, respectively. The terms $E_{1t}$, $E_{2t}$ and $E_{3t}$ are the noise variables that follow a multivariate Gaussian distribution $\mathcal{N}(\textbf{0},\Sigma)$. 

\begin{figure*}[!ht]
		\centering
		\begin{subfigure}{0.32\textwidth}
			\centering
			\includegraphics[width=0.9\linewidth]{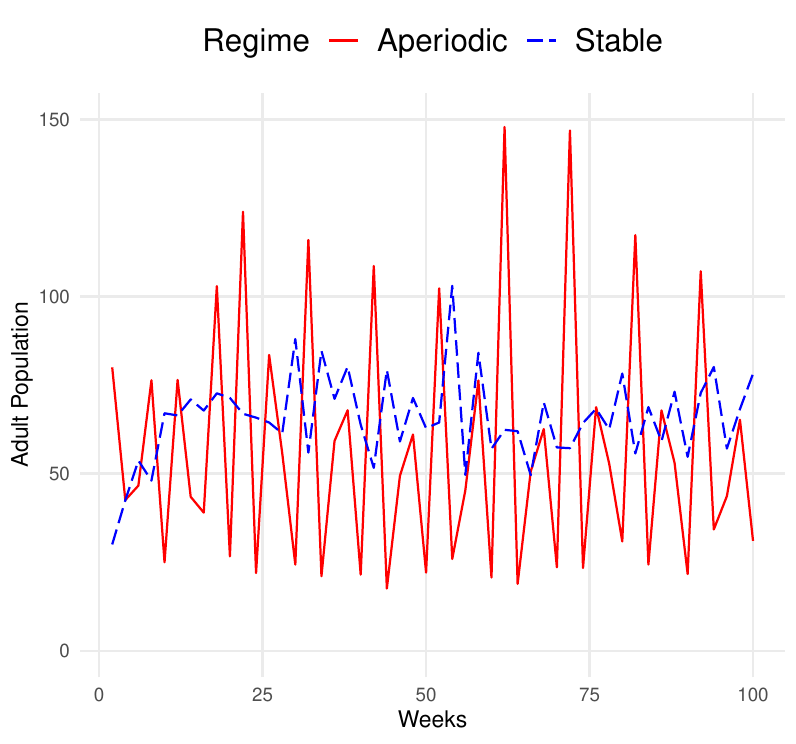}
			\caption{Tribolium adult population of aperiodic and stable regimes}
			\label{subfig:beetlepop}
		\end{subfigure}
		\begin{subfigure}{0.32\textwidth}
			\centering
			\includegraphics[width=0.9\linewidth]{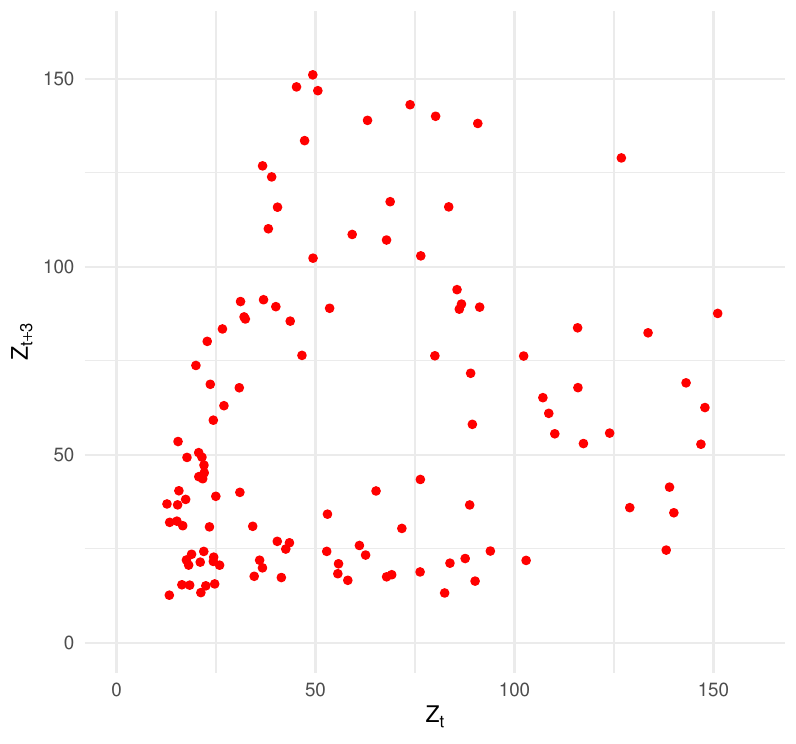}
			\caption{Embedded point cloud data of aperiodic oscillation population }
			\label{subfig:beetleaperiodic}
		\end{subfigure}
		\begin{subfigure}{0.32\textwidth}
			\centering
			\includegraphics[width=0.9\linewidth]{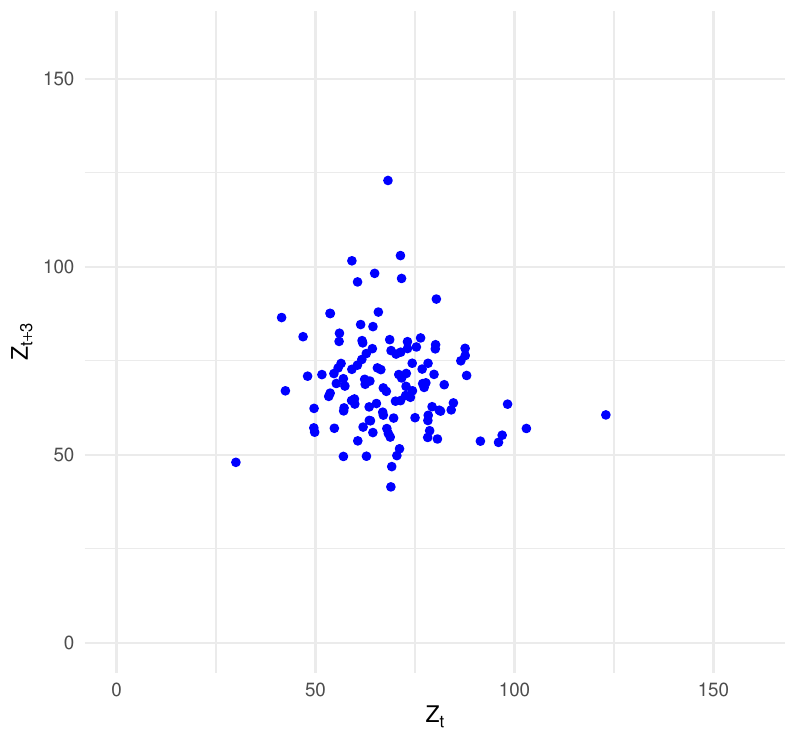}
			\caption{Embedded point cloud data of stable equilibrium population}
			\label{subfig:beetlestable}
		\end{subfigure}
		\caption{Simulated Tribolium populations of aperiodic and stable regimes and their reconstructed point data in $\mathbb{R}^2$ using $\tau=3$.}
		\label{fig:beetlepop}
	\end{figure*}

For the model parameters, we set $b=7.48$, $c_{ea}=0.009$, $c_{pa}=0.004$, $c_{el}=0.012$, $\mu_p=0$, and $\mu_l=0.267$, as used in \cite{costantino1995experimentally}. Also, we use $\Sigma=0.1^2I_3$, where $I_3$ is the identity matrix of size 3.
We compare the populations of two different regimes: 1) stable equilibrium at $\mu_a=0.73$ and 2) aperiodic oscillation at $\mu_a=0.96$. The oscillation pattern could cause greater harm because sudden overpopulation could be difficult to predict \citep{pereira2015persistent}. Figure~\ref{subfig:beetlepop} shows 100 weeks of simulated population data from the two regimes.
We randomly generate Tribolium population data for a total of 240 weeks, recorded every two weeks.

Topological features of reconstructed time series can detect differences.
\cite{pereira2015persistent} show that applying K-means clustering to the raw Tribolium population data is not successful in identifying the stable and aperiodic regimes.
We reconstruct the Tribolium adult population data using Takens' embedding theorem \citep{takens1981detecting}. Let the time series be $Z=\{z_1,z_2,\cdots,z_n\}$. For a given embedding dimension $d$ and time delay $\tau$, $Z$ can be reconstructed as $Z^{\text{Takens}}=[Z_1\; Z_2\; \cdots Z_N ]^T$, where $Z_i=\{ z_i,z_{i+\tau},\cdots,z_{i+(d-1)\tau} \} $. As a result, the time series $Z$ is reconstructed as point cloud data $Z^{\text{Takens}}$ in $\mathbb{R}^d$. Figures~\ref{subfig:beetleaperiodic} and~\ref{subfig:beetlestable} show the reconstructed time series of the aperiodic oscillation and stable equilibrium regimes of Figure~\ref{subfig:beetlepop} when $d=2$ and $\tau=3$. 
We construct the Rips complexes using the point cloud data $Z^{\text{Takens}}$ and compute persistent homology.

We conduct hypothesis tests to examine power and false positive rate. To investigate power, 40 Tribolium adult populations are generated; 20 from the stable equilibrium region and 20 from the aperiodic oscillation regime. The hypothesis test is conducted for the 40 populations. To compute the false positive rate, 40 populations are simulated from the aperiodic oscillation regime. The generated populations are randomly assigned to two groups of equal size and the hypothesis test is conducted between two groups. Similarly, we generate 40 populations from the stable equilibrium regime and conduct the hypothesis test.
We repeat this procedure 100 times and a total of $3\times100\times40=12,000$ populations are generated. 
The differences between two groups are compared with a loop in a reconstructed point cloud, so dimension-one persistent homology results are used for hypothesis testing.
In the two-stage test, the persistence images of size 40 by 40 are generated using the Gaussian smoothing function with $h=3$ and the arctangent weight. In the filtering and testing stages, we use $C=80\%$ and the BH procedure.
For additional comparison, we also conduct the permutation test using the dynamic time warping (DTW) distance between time series \citep{giorgino2009computing}.
For PD, PL, and DTW, $N_P=100$, $N_P=500$, and $N_P=500$ are used, respectively.

			\begin{table*}[!t]
		\caption{Power and false positive rate of four hypothesis testing methods of Tribolium beetle population data.}
		\tabcolsep=0pt
\begin{tabular*}{\textwidth}{@{\extracolsep{\fill}}lccc@{\extracolsep{\fill}}}
\hline
          &      Stable vs. Aperiodic & Aperiodic vs. Aperiodic  & Stable vs. Stable\\
          & (Power) & (False Positive Rate) &  (False Positive Rate) \\\hline
Two-stage & 1.00                                                         & 0.01                                                                       & 0.00                                                                       \\ 
PD        &     1.00                                                         &        0.04                                                                    &            0.06                                                                \\ 
PL        & 1.00                                                         & 0.02                                                                       & 0.07                                                                       \\ 
Kernel    &   1.00                                                           &         0.02                                                                   &     0.06                                                                       \\ 
DTW    &   1.00                                                           &         0.03                                                                   &     0.03     \\                      \hline
\end{tabular*}
	\label{tb:beetle}
	\end{table*}

Table~\ref{tb:beetle} summarizes the false positive rates and powers of three scenarios. All hypothesis testing methods achieve high powers and low false positive rates. The hypothesis testing results suggest that the topological features of the embedded Tribolium adult population time series data can identify two regimes. Figure~S9 in the supplementary material shows p-values of the two-stage hypothesis test between the stable and aperiodic regimes. The results suggest that the differences between the two regimes are due to 1) the number of small-sized loops and 2) the number of loops whose diameters are about 25.

	\section{Application Study}
	\label{sec:app}

	\subsection{Sand Pack Image Data Analysis}
	\label{subsec:real}
	
\begin{figure}[!t]
		\centering
		\begin{subfigure}{0.28\textwidth}
			\centering
			\includegraphics[width=0.9\linewidth]{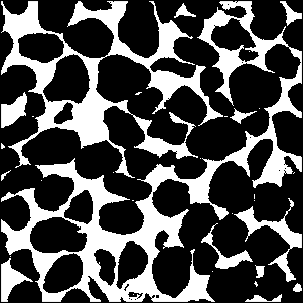}
			\caption{Unground silica F42}
			\label{subfig:f42}
		\end{subfigure}
		\begin{subfigure}{0.28\textwidth}
			\centering
			\includegraphics[width=0.9\linewidth]{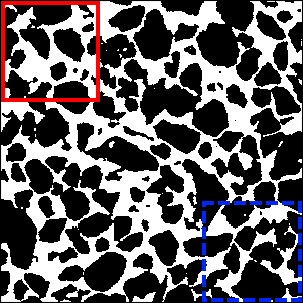}
			\caption{Levenseat sand LV60}
			\label{subfig:lv60}
		\end{subfigure}
		\caption{2D slice images of two sand packs. Pores and grains are drawn as white and black pixels. The red solid and the blue dotted box areas are two of the subregions that show large variabilities within the same rock image.}
		\label{fig:rock}
	\end{figure}

	As an example of the use of persistent homology for the analysis of real imaging data, we analyze two sand packs: F42 (unground silica, US Silica Company) and LV60 (Levenseat sand, WBB Minerals, UK). 
	These sand pack datasets are obtained by nuclear magnetic resonance scans and Micro-CT imaging by \cite{Talabi2009}. 
	Each sand pack dataset includes two-samples: F42B and F42C and LV60A and LV60C.
	All sand pack images have $300^3$ voxels with resolution 10.002 $\mu m$.
	Table~S2 in the supplementary material summarizes the sand pack data and Figure~\ref{fig:rock} shows their 2D slice images.

	Both rock types have similar porosity (volume of pores divided by total volume), but different grain surface area.   
	In Figure~\ref{fig:rock}, we see that the unground silica sand pack has larger-sized and circular-shaped grains compared to the Levenseat sand pack. However, it is difficult to compare structural and connectivity differences from the summarized properties and the 3D images themselves. 
	
	We compute persistent homology to extract structural and connectivity information of sand pack data. 
	First, we take 27 subregion images of size $100^3$ from each sand pack sample.  
	The persistence diagrams are converted into persistence images using the arctangent weight.

	We conduct hypothesis tests on three sets of data: between unground silica (F42B and F42C), between Levenseat sand (LV60A and LV60C), and between unground silica and Levenseat sand (F42B and LV60A). 
	For PD, we use $N_P=500$ permutations for dimension zero and two, and $N_P=100$ for dimension one due to the computation time of pairwise distances. Also, we use 1-Wasserstein distance as a pairwise distance. 
	For two-stage test, persistence images are converted using the Gaussian smoothing function with $h=1.5$, the arctangent weight, and filtering threshold $C=50$\%. 
	For PL, we use $N_P=1,000$. For Kernel, the arctangent weight, the Gaussian kernel of $h=1.5$, and 1,000 bootstrap samples are used.

\begin{figure*}[!ht]
		\centering
		{\hspace{0.6in} F42B and F42C \hspace{0.7in} LV60A and LV60C \hspace{0.7in} F42B and LV60A}\\
		{\rotatebox{90}{Dimension-zero}}
		\begin{subfigure}{0.315\textwidth}
			\centering
			\includegraphics[width=0.95\linewidth]{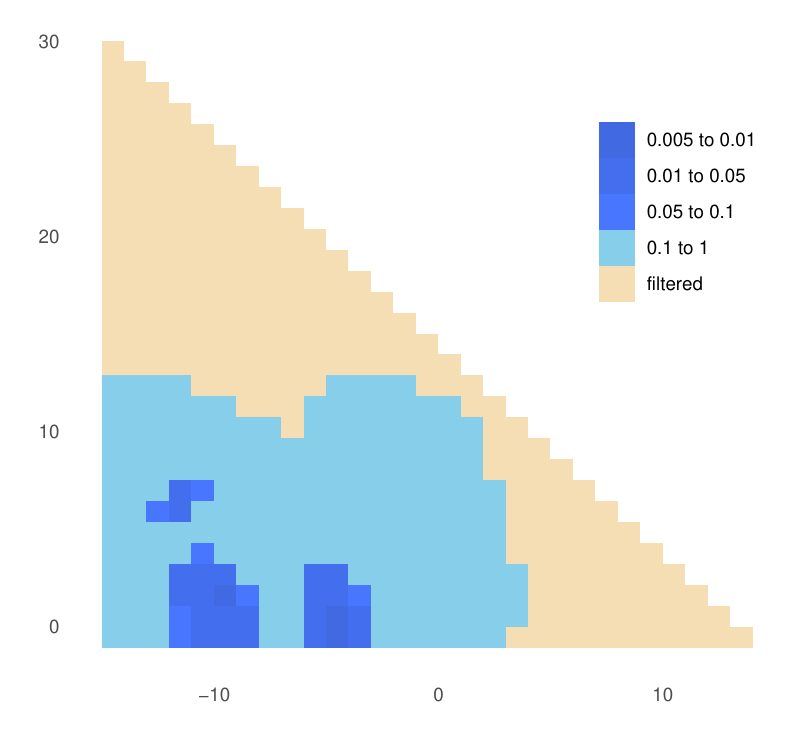}
			\label{subfig:rockF42dim0}
		\end{subfigure}
		\begin{subfigure}{0.315\textwidth}
			\centering
			\includegraphics[width=0.95\linewidth]{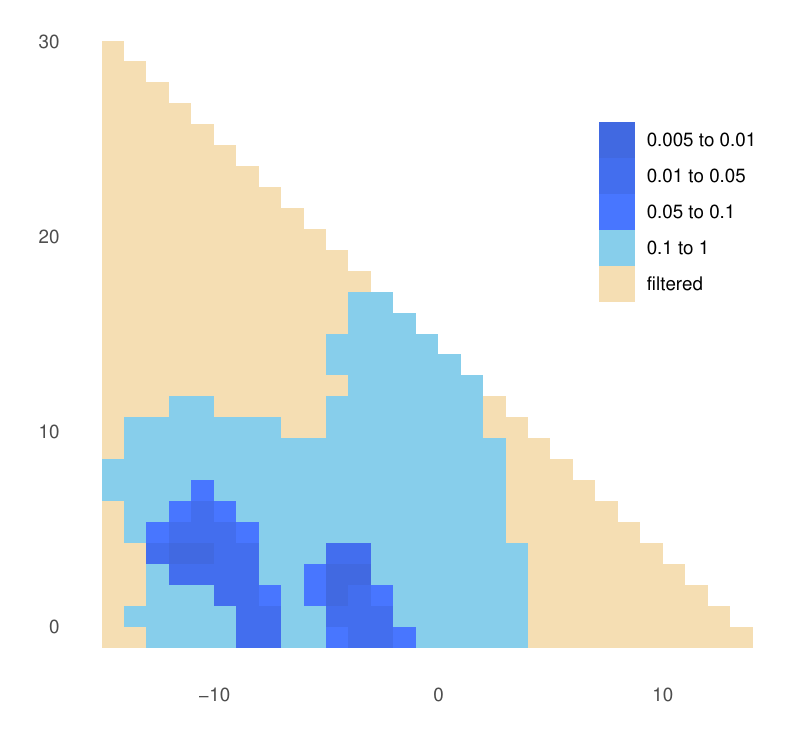}
			\label{subfig:rockLV60dim0}
		\end{subfigure}
		\begin{subfigure}{0.315\textwidth}
			\centering
			\includegraphics[width=0.95\linewidth]{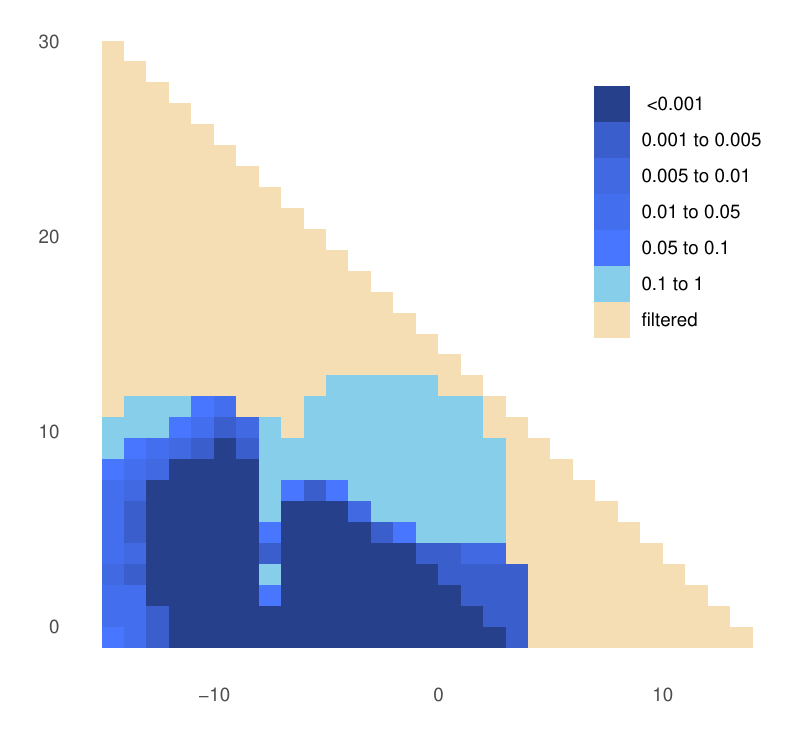}
			\label{subfig:rockFLVdim0}
		\end{subfigure}
		
		{\rotatebox{90}{Dimension-one}}
		\begin{subfigure}{0.315\textwidth}
			\centering
			\includegraphics[width=0.95\linewidth]{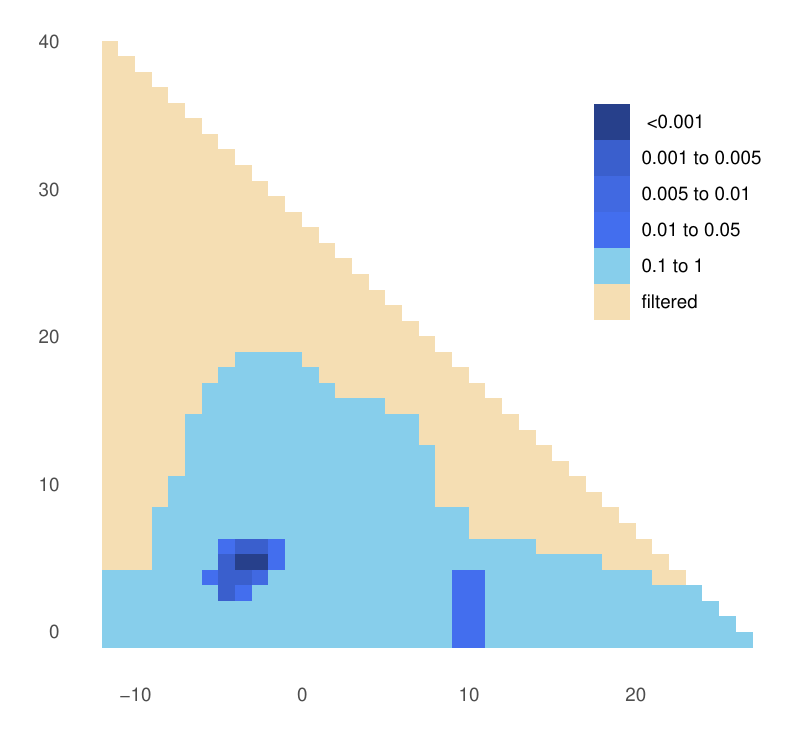}
			\label{subfig:rockF42dim1}
		\end{subfigure}
		\begin{subfigure}{0.315\textwidth}
			\centering
			\includegraphics[width=0.95\linewidth]{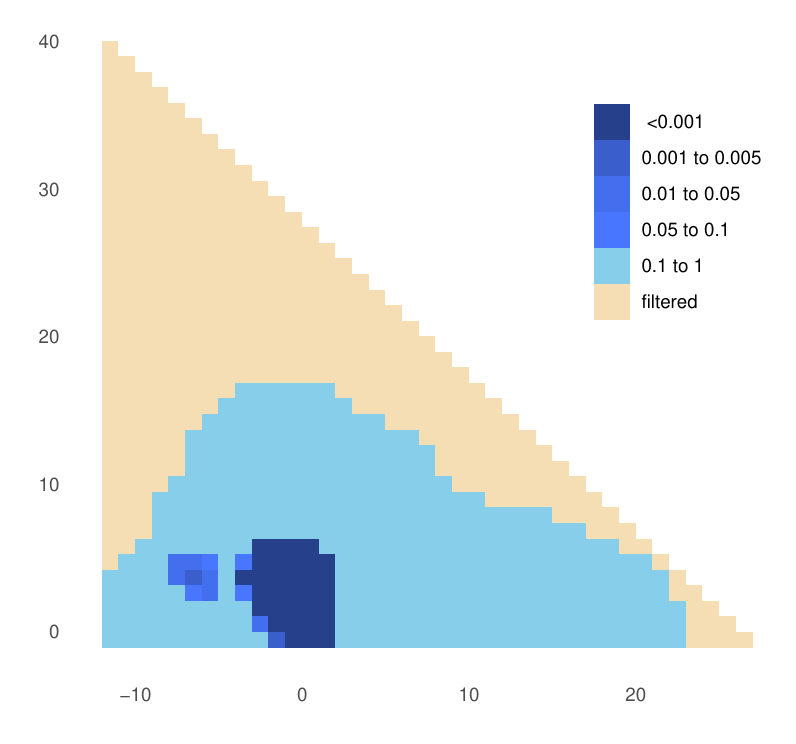}
			\label{subfig:rockLV60dim1}
		\end{subfigure}
		\begin{subfigure}{0.315\textwidth}
			\centering
			\includegraphics[width=0.95\linewidth]{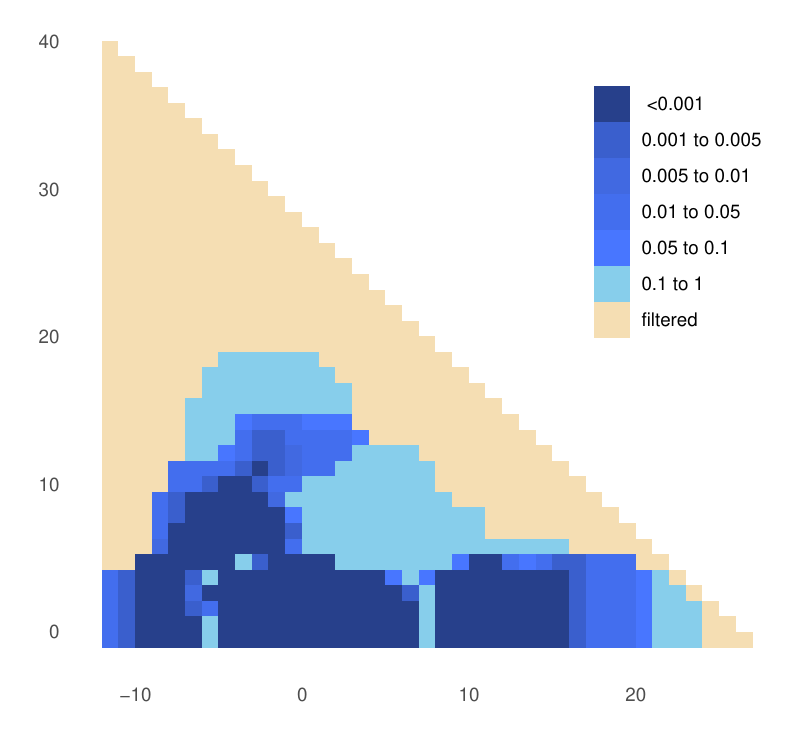}
			\label{subfig:rockFLVdim1}
		\end{subfigure}
		
		{\rotatebox{90}{Dimension-two}}
		\begin{subfigure}{0.315\textwidth}
			\centering
			\includegraphics[width=0.95\linewidth]{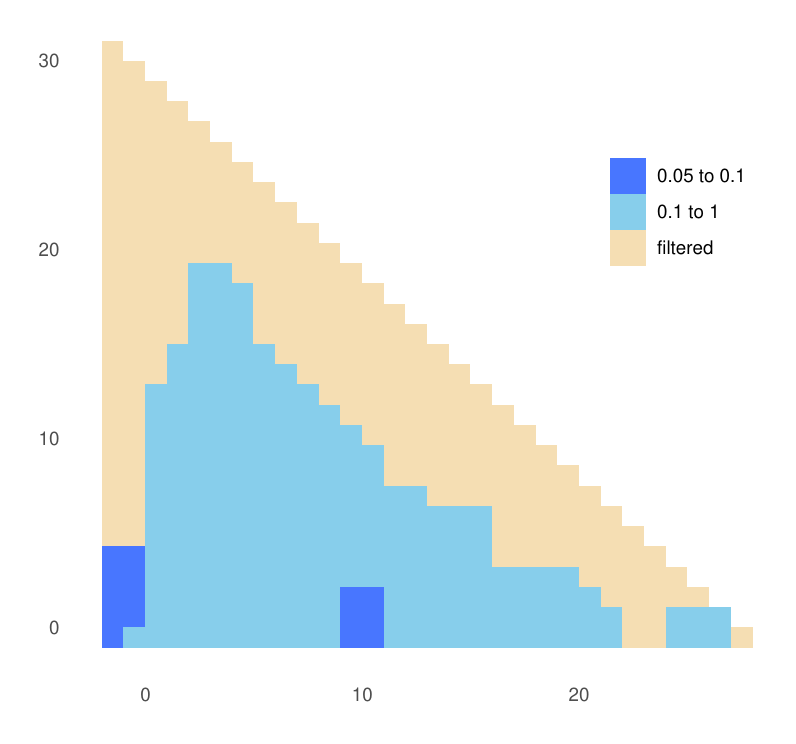}
			\label{subfig:rockF42dim2}
		\end{subfigure}
		\begin{subfigure}{0.315\textwidth}
			\centering
			\includegraphics[width=0.95\linewidth]{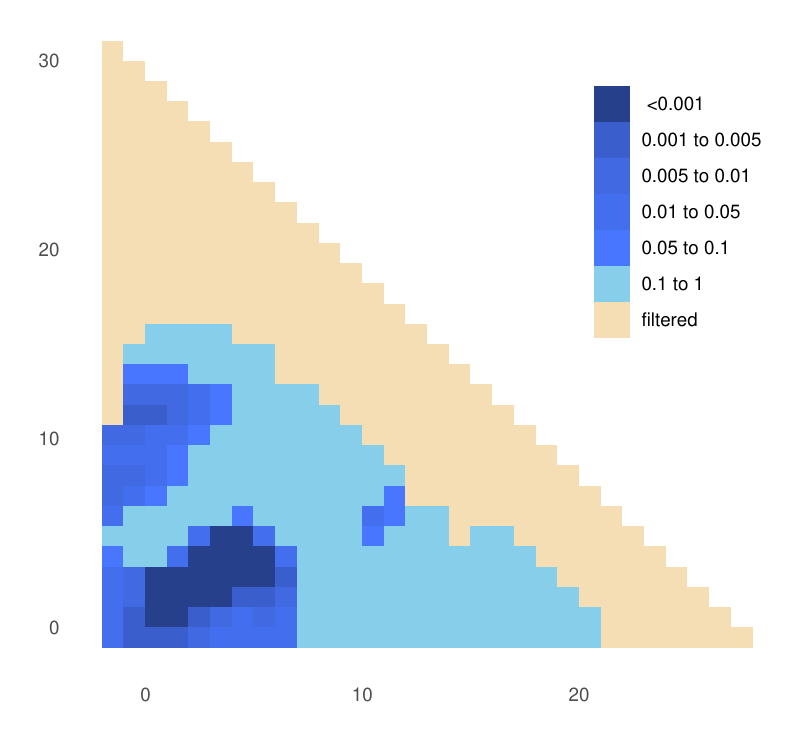}
			\label{subfig:rockLV60dim2}
		\end{subfigure}
		\begin{subfigure}{0.315\textwidth}
			\centering
			\includegraphics[width=0.95\linewidth]{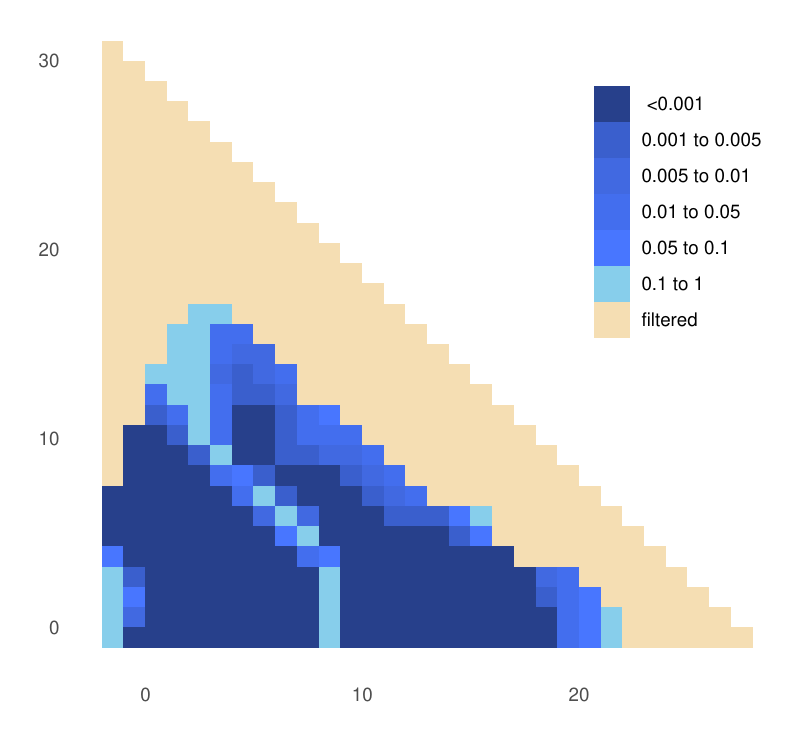}
			\label{subfig:rockFLVdim2}
		\end{subfigure}
		\caption{p-values of two-stage hypothesis tests between F42 (first column), between LV60 (second column), and between F42 and LV60 (third column) for dimension-zero (first row), dimension-one (second row), dimension-two (third row) persistence images. 
		}
		\label{fig:rock.result}
	\end{figure*}

	Table~S3 in the supplementary material shows the hypothesis testing results. The four tests yield similar testing results; p-values are small in most cases. 
	Figure~\ref{fig:rock.result} presents the p-values of two-stage hypothesis tests that show which structural and connectivity differences exist between the sand pack images. 

	For example, the hypothesis test using the dimension-zero persistence images between F42 rocks (top left of Figure~\ref{fig:rock.result}) implies that connectivity of grains of size 10 and 5 differs the most. The dimension-one test result between F42 rocks (middle left of Figure~\ref{fig:rock.result}) suggests that the number of pores of size 5 differs the most.
	Also, the tests between F42 and LV60 sand packs (right of Figure~\ref{fig:rock.result})  have larger areas of small p-values than the same type of sand packs. 
	This implies that various types of topological features account for differences between the two types of rocks.

	We note that the small p-values for both methods in the same type of rocks might be due to the small subregion size. 
	The subregion size $100^3$ is relatively small to represent the overall rock sample structure. 
	The sampled subregions may not have similar structures and connectivity, even when they are taken from the same rock sample. 
	For example, we observe that subregions from the same rock sample may have large variabilities themselves in Figure~\ref{subfig:lv60}; the subregion of size 100 by 100 in the top-left corner (in the red solid line box) has smaller sized and more sparse grains than the bottom-right corner subregion (in the blue dashed line box).

	\subsection{Musical Instrument Sound Data}
	\label{subsec:sound}
	
\begin{figure}[!ht]
		\centering
		\begin{subfigure}{0.32\textwidth}
			\centering
			\includegraphics[width=1\linewidth]{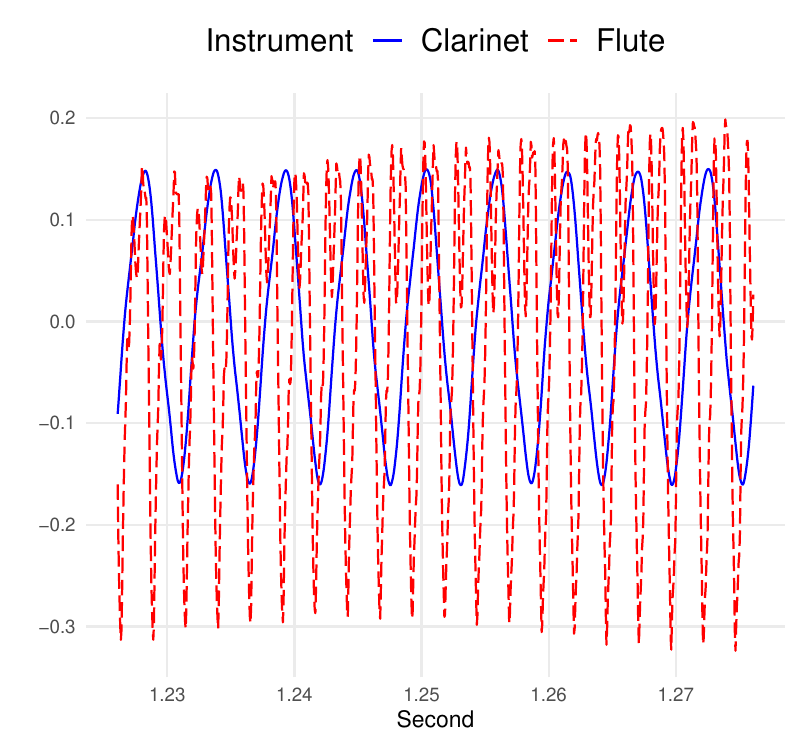}
			\caption{Clarinet and flute sound waves}
			\label{subfig:insttime}
		\end{subfigure}
		\begin{subfigure}{0.32\textwidth}
			\centering
			\includegraphics[width=1\linewidth]{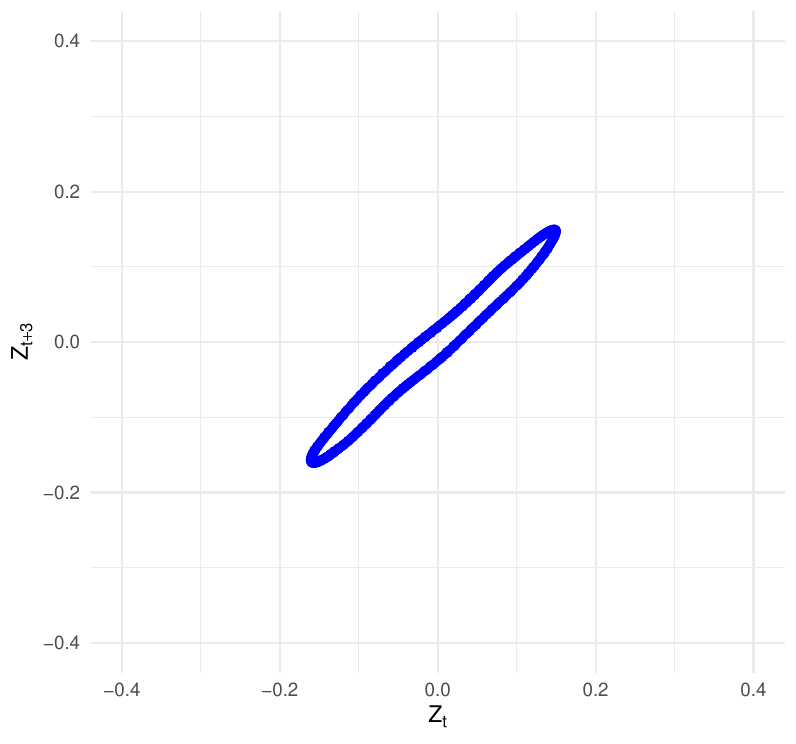}
			\caption{Embedded clarinet sound data}			\label{subfig:instclarinet}
		\end{subfigure}
		\begin{subfigure}{0.32\textwidth}
			\centering
			\includegraphics[width=1\linewidth]{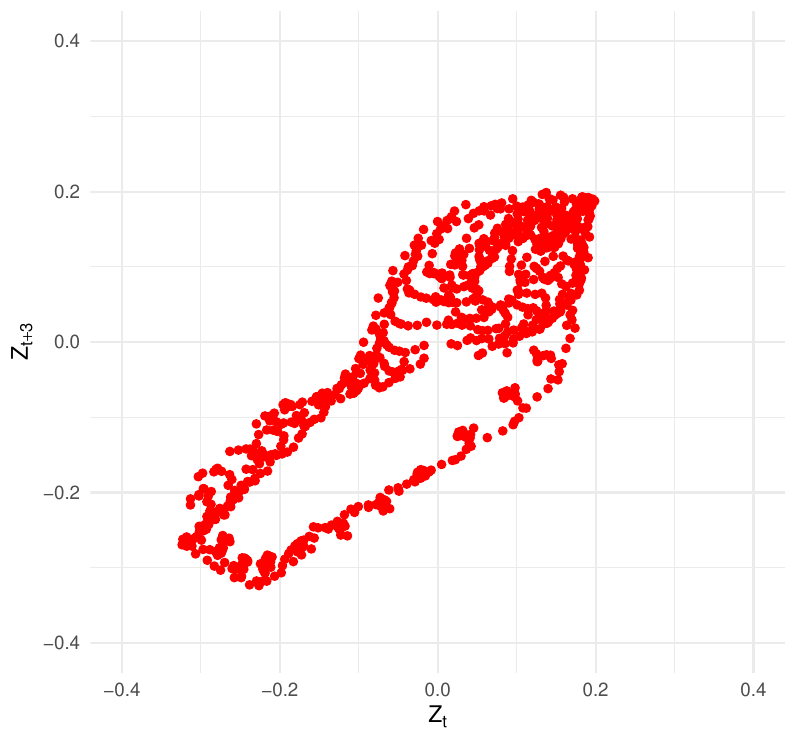}
			\caption{Embedded flute sound data}
			\label{subfig:instflute}
		\end{subfigure}
		\begin{subfigure}{0.32\textwidth}
			\centering
			\includegraphics[width=1\linewidth]{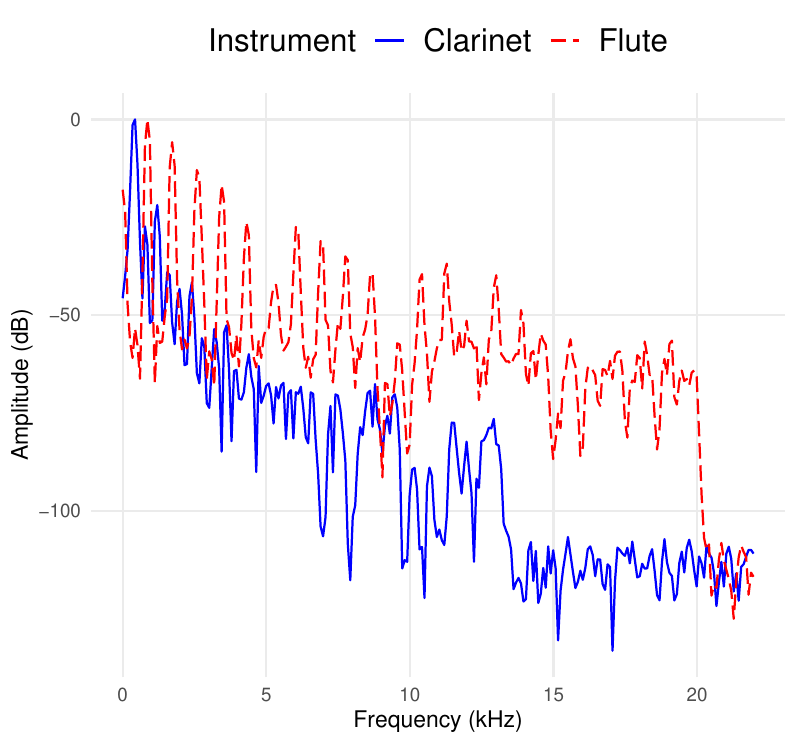}
			\caption{Mean frequency spectrum}
			\label{subfig:instfrequency}
		\end{subfigure}
		\begin{subfigure}{0.32\textwidth}
			\centering
			\includegraphics[width=1\linewidth]{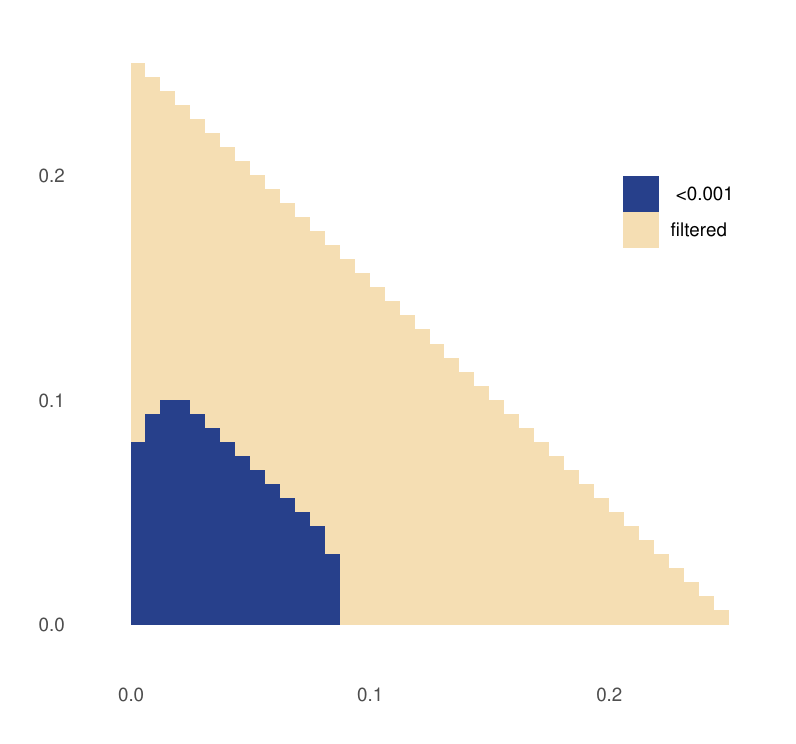}
			\caption{p-values of two-stage test}
			\label{subfig:instpval}
		\end{subfigure}
		\caption{Sound waves of clarinet and flute, embedded sound data, mean frequency spectrum, and p-values of two-stage hypothesis test between clarinet and flute sound data.}
		\label{fig:inst}
	\end{figure}

	We conduct hypothesis tests using sound data of two wind instruments, flute and clarinet. The sound data are available in the public repository (\url{https://github.com/MattO-Reilly/TDA-TimeSeriesAnalysis}). For both instruments, note A4 at 44,100 Hz is used. The clarinet and flute sounds are recorded for 4.75 and 1.9 seconds, respectively, with a time unit of about 50 microseconds.
	Figures~\ref{subfig:insttime}, \ref{subfig:instclarinet}, and \ref{subfig:instflute} show about 0.05 seconds sampled sound of the two instruments and the corresponding reconstructed point cloud data. 
	
	Three scenarios are tested: 1) clarinet and flute sound data, 2) between clarinet sounds, and 3) between flute sounds. For each scenario, we sample about 0.05 seconds of sound data (1,000 time units) from each group. For both instruments, the samples are taken after 0.5 seconds when the tones are stabilized. The sampling process is repeated 20 times. The sampled sound data are reconstructed by Takens' embedding as point clouds using $d=2$ and $\tau=3$. From the point cloud data, the Rips complexes are constructed and we compute persistent homology. We compare loops in point clouds and use the dimension-one persistence diagrams in the hypothesis tests. In the two-stage test, persistence images with the Gaussian smoothing function with $h=0.025$ are used. The other parameters are the same as used in Section~\ref{subsec:beetle}. 
	
	Figure~\ref{subfig:instpval} presents the p-values of two-stage hypothesis test between clarinet and flute sound data. The result indicates that the number of small sized one-dimensional holes differs for the reconstructed sound data of the two instruments. The small-sized loops of the embedded point clouds are generated due to the amplitude fluctuations and high-frequency oscillations of the sound waves. For the note A4 data, the flute has a stronger high-frequency sound spectrum than the clarinet. Figure~\ref{subfig:instfrequency} shows the mean frequency spectrum of two instruments. The sound of the flute playing the A4 note has more high-amplitude peaks at high-frequencies than the sound of the clarinet.
	
	The p-values of scenarios 2 and 3, between clarinet sounds and between flute sounds, are larger than that of scenario 1, between different instruments. The minimum p-values of two-stage hypothesis tests between flute sounds and between clarinet sounds are 0.992, and 0.968, respectively.

	\section{Conclusion}
	\label{sec:discussion}
	This paper proposes two-stage hypothesis test that consists of filtering and testing steps for the persistence image. 
	The proposed approach enables better inference by 1) achieving higher power than existing methods while maintaining low false positive rate, 2) providing specific regions on the persistence images that contribute the most to any observed differences, and 3) implementing flexible weights to topological features.
	The simulation studies and real data analysis show that the proposed method performs well by comparing shapes of data represented by topological features.
	
	In general, the computation times of the four hypothesis testing methods are not a significant issue in our experience. The only computational bottleneck is the pairwise distance computations used in \cite{Robins2016}.
	This is because we use the Hungarian algorithm to compute the 1-Wasserstein distance that has $O(n_f^3)$ time complexity. 
	However, the computational cost could be significantly reduced by using the approximation algorithms for the Wasserstein and bottleneck distances \citep{Kerber2017,chen2021approximation}.
	
	The proposed two-stage test is mainly applied to the persistence image in our study. However, it could be extended to other representations of persistence diagrams. For example, when the two-stage test is applied to the two-sample z-test using mean persistence landscapes proposed in \cite{Bubenik2015}, it may reduce the number of landscape points used in the test. We present the two-stage hypothesis test algorithm for the persistence landscape in Algorithm~S2 in the supplementary material. Also, the two-stage hypothesis test can be applied to discretized persistence landscapes and the Fourier features of kernels \citep{rahimi2007random}.
	
	Several interesting developments and future topics still remain. 	
	First, the systematical methods to select the weights and parameters can be explored. 
	The proposed method provides a flexible way to conduct a hypothesis test for persistent topological features by implementing weights and smoothing.
	The results in Figure~\ref{fig:weight} indicate that the weights play an important role in a successful result in revealing the differences between persistence diagrams.
	However, the weight selection process itself is still an unanswered question. 
	Also, selecting the filter threshold $C$ can be an issue. 
	One of the disadvantages of the two-stage hypothesis testing is that the choice of the filtering threshold $C$ could be subjective \citep{Du2014}. We may try a data-driven greedy independent filtering procedure of \cite{Ignatiadis2016} that chooses the threshold that maximizes the number of discoveries among all possible candidates. 
	Second, it might be worth studying the potential dependence structures of persistence images. The dependence structure of pixels in persistence images could be different from dataset to dataset and can change after filtering. Although our simulation study suggests that the multiple testing adjustments work well under the independence assumption, it would be helpful to identify potential dependency under various settings.
 
 \section*{Supplementary Material}
Supplementary material contains supplementary sections and figures.

 \section*{Data Availability}
    The code and data underlying this article are available in the GitHub repository at \url{https://github.com/chulmoon/HT-VecPD}.




\bibliographystyle{asa}
\bibliography{reference}

\end{document}